\documentclass[sigplan,nonacm,balance=false]{acmart}
\usepackage{balance} 
\renewcommand\footnotetextcopyrightpermission[1]{}

\makeatletter
\def\@oddhead{}  % 彻底清空所有单数/奇数页的页眉
\def\@evenhead{} % 彻底清空所有双数/偶数页的页眉
\makeatother
\AtBeginDocument{%
  }

\usepackage{graphicx}
\usepackage{subcaption}
\usepackage{booktabs}
\usepackage{tabularx}
\usepackage{array}
\usepackage{ragged2e}
\usepackage{enumitem}
\usepackage{algorithm}
\usepackage{algpseudocode}
\usepackage{hyperref}
\hypersetup{
    colorlinks=true,
    linkcolor=blue,
    citecolor=blue
}
\usepackage{comment}
\usepackage{placeins}
\usepackage{float}
\usepackage{microtype}
\graphicspath{{./figures/}}
\newcolumntype{L}{>{\RaggedRight\arraybackslash}X}
\newcolumntype{P}[1]{>{\RaggedRight\arraybackslash}p{#1}}

\usepackage{CJKutf8}

\usepackage{xcolor}

\usepackage{xspace}
\newcommand{\sys}{HyperParallel-FSDP\xspace}

\newif\ifshowcomment
\showcommenttrue
\ifshowcomment
  \newcommand{\todo}[1]{\textcolor{red}{[\begin{CJK}{UTF8}{gbsn}TODO:~#1\end{CJK}]}}
\else
  \newcommand{\todo}[1]{}
\fi
\usepackage{tikz}
\usetikzlibrary{positioning,arrows.meta,fit,backgrounds,calc,shapes.geometric}
\definecolor{hpblue}{RGB}{36,87,153}
\definecolor{hpgreen}{RGB}{27,122,75}
\definecolor{hporange}{RGB}{196,110,22}
\definecolor{hpred}{RGB}{176,48,40}
\definecolor{hpgray}{RGB}{85,85,85}
\tikzset{
  every picture/.style={font=\footnotesize},
  fbox/.style={draw=#1!80, rounded corners=1.5pt, fill=#1!6, align=center,
               inner sep=2.5pt},
  fbox/.default=hpblue,
  bad/.style={fbox=hpred},
  good/.style={fbox=hpgreen},
  warn/.style={fbox=hporange},
  ghost/.style={fbox=hpgray},
  flow/.style={-{Stealth[length=2mm]}, thick, draw=hpgray!90},
  gflow/.style={-{Stealth[length=2mm]}, thick, draw=hpgreen},
  oflow/.style={-{Stealth[length=2mm]}, thick, draw=hporange},
  fnote/.style={font=\scriptsize, text=hpgray, align=center},
  flabel/.style={font=\scriptsize\itshape, text=hpgray},
}
\usepackage{eso-pic}

\newcommand{\addAIStatement}{%
  \AddToShipoutPictureFG*{%
    \AtPageLowerLeft{%
      \raisebox{13mm}{%
        \makebox[\paperwidth][c]{%
          \parbox{0.82\paperwidth}{%
            \centering
            \scriptsize\itshape
            Generative AI tools were used to assist with manuscript drafting,
            organization, and language refinement. All AI-assisted content was
            reviewed and verified by the authors, who take full responsibility
            for the final manuscript.
          }%
        }%
      }%
    }%
  }%
}

\makeatletter
\renewcommand{\country}[1]{%
  \global\@ACM@countrypresenttrue
  \unskip\ignorespaces
}
\makeatother

\begin{document}

\title{\sys: Topology-Aware Fully Sharded Training with Layout-Driven Muon on Ascend SuperPods}

\author{Mo Sun}
\authornotemark[1]
\affiliation{%
  \institution{Zhejiang University}\country{\mbox{}}
}

\author{Yifan Yao}
\authornote{Contributes equally.}
\affiliation{%
  \institution{Huawei Technologies Co., Ltd}\country{\mbox{}}
}

\author{Yanwei Liu}
\affiliation{%
  \institution{Huawei Technologies Co., Ltd}\country{\mbox{}}
}

\author{Luobin Liu}
\affiliation{%
  \institution{Huawei Technologies Co., Ltd}\country{\mbox{}}
}

\author{Zhenzhang Yang}
\affiliation{%
  \institution{Huawei Technologies Co., Ltd}\country{\mbox{}}
}

\author{Kaisheng Wang}
\affiliation{%
  \institution{Huawei Technologies Co., Ltd}\country{\mbox{}}
}

\author{Xiangyu Meng}
\affiliation{%
  \institution{Huawei Technologies Co., Ltd}\country{\mbox{}}
}

\author{Chen Li}
\affiliation{%
  \institution{Huawei Technologies Co., Ltd}\country{\mbox{}}
}

\author{Xizheng Pang}
\affiliation{%
  \institution{Huawei Technologies Co., Ltd}\country{\mbox{}}
}

\author{Huilan Li}
\affiliation{%
  \institution{Huawei Technologies Co., Ltd}\country{\mbox{}}
}

\author{Xinglei Xu}
\affiliation{%
  \institution{Huawei Technologies Co., Ltd}\country{\mbox{}}
}

\author{Yushi Cui}
\affiliation{%
  \institution{Huawei Technologies Co., Ltd}\country{\mbox{}}
}

\author{Xinyao Lin}
\affiliation{%
  \institution{Zhejiang University}\country{\mbox{}}
}

\author{Kaiqi Chen}
\affiliation{%
  \institution{Zhejiang University}\country{\mbox{}}
}

\author{Jie Zhang}
\affiliation{%
  \institution{Zhejiang University}\country{\mbox{}}
}

\author{Zeke Wang}
\affiliation{%
  \institution{Zhejiang University}\country{\mbox{}}
}

\author{Teng Su}
\affiliation{%
  \institution{Huawei Technologies Co., Ltd}\country{\mbox{}}
}

% \affiliation{%
%   \vspace{1em}
%   \institution{\textsuperscript{1}~Huawei Technologies Co., Ltd, China\\[2pt]
%                \textsuperscript{2}~Zhejiang University, China}
%   \country{}
% }

\begin{abstract}
% Abstract (rewritten per the 2026-09-02 integration directive, sec. 6.2;
% experimental scope updated 2026-09-03 to the delivered data, incl. the
% Muon comparison reports): four parts
% -- problem, key insight, three designs, current experiment scope.

Declarative SPMD programming, in which a sharding description attached to
tensors drives all distributed execution, decouples parallelization from
model code. The evaluated eager stack built on PyTorch's native
distributed tensor, however, dispatches every operator \emph{below} the
autograd engine, paying per-operator dispatch and metadata-handling cost
on each training step, and offers no low-cost end-to-end validation path
for the same plan used in production. FSDP communication and distributed
Muon expose two further mismatches on two-tier supernode networks, whose
intra- and inter-supernode interconnects differ by an order of magnitude:
the evaluated fully-sharded configuration materializes parameters through
explicit packing and unpacking passes, and the increasingly adopted Muon
optimizer orthogonalizes whole matrices, which conflicts with parameter
sharding.

Our key insight is that distributed tensors only need to describe sharding
semantics at the tensor API boundary \emph{above} autograd (i.e., the
function-level interception point before the autograd engine records
operations), so that differentiation and kernel execution always observe
plain tensors. We present \sys, a training system built on this insight
with three designs:
(1)~\emph{Dual-Mode Distributed-Tensor Execution}, where we propose one declarative
sharding plan to drive both a production mode that resolves layouts once at
plan-application time with zero steady-state per-operator DTensor dispatch
overhead, and a validation mode that propagates layout metadata
end-to-end with fail-fast checking, closed by a gradient-equivalence
harness.
(2)~\emph{Topology-Aware Fully-Sharded Data Parallelism}, where we propose zero-copy
per-parameter collectives for the dominant leading-dimension-sharded
layouts within a supernode, fused reduction across supernodes, and a
four-step cross-layer backward pipeline that keeps the backward path free
of layer-level waits on slow links.
(3)~\emph{Layout-Driven Distributed Muon}, including communication groups derived
from sharding semantics, deduplication of orthogonalization across shard
and replica domains, and shape-fused batched Newton--Schulz iterations.

We evaluate \sys on Atlas 900 A3 SuperPoD from 16 dies 
to 384 physical cards (768 ranks).
% production and validation modes execute identical device work,
% framework-recorded peak memory, and loss trajectories over the evaluated
% horizon; 
The topology-aware fully-sharded layers sustain a 505B-parameter MoE at
421k tokens/s at 768 ranks with FSDP communication occupying 2.9\% of
step time, which reduces mean step time by 29.7\% against the evaluated 
PyTorch FSDP2 configuration at 16 dies and by 25.5\% % (1.34$\times$) 
against Megatron DDP with distributed optimizer at 128 ranks, 
while per-step losses track the baseline at Pearson
$r>0.999997$ over 1{,}000 steps.
The layout-driven distributed Muon reduces profiler step
time by 5.4\textasciitilde 16.0\% against DMuon, MatrixFSDP, and TorchTitan
FlexShard/DistMuon integrations at 16 NPUs. %---comparisons that also
% surface an owner-side memory imbalance (an out-of-memory at a 40-layer
% configuration that \sys trains) and a recompute-exposed
% gradient-lifetime bug in the owner-resident baselines, neither of which
% the shard-native design has.
\sys is open source at \url{https://atomgit.com/mindspore/hyper-parallel}.

\end{abstract}

% \settopmatter{printfolios=true}

\addAIStatement
\maketitle

% \begingroup
% \renewcommand\thefootnote{$\dagger$}
% \footnotetext{Contributes equally.}
% \endgroup

\enlargethispage{2\baselineskip} % 确保脚注留在第一页底部
\pagestyle{plain}

\section{Introduction}
\label{sect:intro}

Large language model training has moved to clusters of
$10^3$\textasciitilde $10^4$ accelerator dies organized as supernodes, and at this scale
training efficiency is decided by how communication is organized, not by
raw compute~\cite{megatronlm,megascale}. The hardware premise has changed
with it: modern NPU supernodes connect hundreds of dies through a
high-bandwidth fabric, while inter-supernode traffic traverses a
conventional datacenter network with roughly an order of magnitude lower
bandwidth, producing
a pronounced two-tier network hierarchy~\cite{cloudmatrix384}.

The programming models in use today sit at two poles, neither satisfactory
on this hardware. Megatron-class frameworks deliver high performance, but
their parallelization is deeply coupled with model code. For instance, tensor, pipeline,
and context parallelism are hand-wired into the model definition, together
with custom checkpointing, which is unfriendly to LLM algorithm
developers~\cite{megatronlm}. Declarative SPMD programming, exemplified by
TorchTitan and NeMo-AutoModel, offers a compelling advantage: the model is
decoupled from the parallel strategy, which is expressed as data over a
device mesh~\cite{torchtitan,nemoautomodel}. Yet adoption remains limited in the production stacks we study, and the
existing realizations explain why: built on PyTorch's
native distributed tensor, they perform operator-level dispatch \emph{below}
the autograd engine and pay per-operator dispatch and metadata-handling
overhead for the whole training
run % ---cached propagation fast paths and version-dependent Python/C++ splits
% notwithstanding---
and expose no separate validation mode that checks the
same static plan used by production, leaving manual debugging as the
fallback~\cite{pytorch2,pytorchdtensor}. We argue that performance-versus-usability
trade-off is removable.

FSDP communication and distributed Muon remain mismatched to
supernodes. First, the evaluated PyTorch 2.9 FSDP2 configuration communicates
in fused units surrounded by explicit packing and
unpacking passes~\cite{pytorchfsdp,pytorchfsdp2}. This fusion amortizes message-launch
overhead on generic Ethernet; in communication-rich supernode
configurations, host-side fusion alone may provide limited end-to-end
benefit. Second, the increasingly widespread Muon
optimizer that is validated at frontier scale by Kimi K2's MuonClip, Kimi K3,
DeepSeek-V4, and the GLM series~\cite{muon,moonlight,kimik2,kimik3,deepseekv4,glm45} %, and an increasingly adopted alternative to AdamW---
orthogonalizes \emph{whole} matrices, which conflicts with sharded parameter layouts.
To bridge this gap, existing distributed realizations restore whole-matrix optimizer inputs
through various mechanisms, including bucketed ZeRO-1-style
assignment, explicit per-parameter owner routing, owner-shaped ZeRO-3
placement, strategy-specific static partitioning, and storage-to-compute
resharding~\cite{moonlight,dmuon,matrixfsdp,canzona,flexshard}. In contrast, \sys's
realization differs in how its communication domains are derived and
bounded
(\S\ref{sect:related}): communication domains and deduplication follow
the distribution semantics of general multi-dimensional meshes rather
than a fixed per-strategy assignment; parameters whose matrix plane is
unsharded require no Muon-specific optimizer communication over the
expert-sharding axis;
and replica-domain deduplication is confined to topology-aligned
subgroups (aligned with machine boundaries under the configured rank
mapping) with deterministic recomputation across them, so that under
the canonical mesh-to-topology mapping (Figure~\ref{fig:topo}) the
evaluated configuration sends no optimizer traffic over the slow
interconnect tier.

In this paper, we present \sys, a training system that resolves the
challenges with one decision and two mechanisms built on it.
(1)~\emph{Dual-Mode Distributed-Tensor Execution}
(\S\ref{sect:dtensor}): the enabling decision is to intercept at the
tensor API layer \emph{above} autograd rather than at the operator
dispatch layer below it, so a single declarative sharding plan drives both
a production mode free of steady-state per-operator DTensor dispatch and a
validation mode with fail-fast layout-contract checking, closed by a
gradient-equivalence harness.
(2)~\emph{Topology-Aware Fully-Sharded Data Parallelism}
(\S\ref{sect:fsdp}): per-parameter collectives that are copy-free for the
dominant layouts within a supernode, fused reduction across supernodes,
and a four-step backward pipeline whose layer-level path contains no wait
on slow communication, where residual work is settled once at the end of the
backward pass.
(3)~\emph{Layout-Driven Distributed Muon} (\S\ref{sect:muon}):
communication domains and two-level deduplication derived automatically
from distribution semantics, so redundant orthogonalization is eliminated
within each shard group and each topology-aligned replica subgroup on
general multi-dimensional process meshes.

We evaluate \sys{} on Atlas 900 A3 SuperPoD from 16 dies to 768 ranks, through dual-mode equivalence and host-overhead measurements
(\S\ref{sec:eval-dtensor}), two whole-system fully-sharded
comparisons. Compared to PyTorch FSDP2 at 16 dies, \sys{} reduces mean step time by 29.7\%; compared to Megatron DDP with distributed optimizer at 128
ranks, \sys{} reduces by 25.5\% while per-step losses track the baseline
at Pearson $r>0.999997$ over 1{,}000 steps. Also, a 505B-parameter MoE
scale demonstration at 384 physical cards (\S\ref{sec:eval-fsdp}), and
distributed Muon comparisons against DMuon, MatrixFSDP, and TorchTitan
FlexShard/DistMuon integrations at 16 NPUs, where profiler step time
drops by 5.4\textasciitilde 16.0\% and the owner-resident baselines surface a memory
imbalance and a recompute-exposed gradient-lifetime bug that the
shard-native design avoids (\S\ref{sec:eval-muon}).

In summary, this paper makes the following contributions:
\begin{itemize}
    \item A dual-mode distributed-tensor abstraction built on pre-autograd,
    API-level interception, including zero steady-state per-operator DTensor
    dispatch in production mode
    and fail-fast verification in validation mode, together with a declarative
    sharding planner that derives strategies automatically from model
    structure (\S\ref{sect:dtensor}).
    \item A topology-aware fully-sharded layer combining zero-copy
    per-parameter collectives for eligible leading-dimension layouts, fused
    cross-supernode reduction over per-step buffer views, and a four-step
    backward pipeline with single-point settlement (\S\ref{sect:fsdp}).
    \item A distributed Muon whose communication domains and two-level
    deduplication are derived automatically from distribution semantics,
    exposed to the algorithm developer as a local-tensor interface backed
    by an internally planned global execution, with shape-fused batched
    orthogonalization and pipelining
    (\S\ref{sect:muon}).
    \item A multi-scale evaluation on Atlas 900 A3 SuperPoD (16 dies to 384
    physical cards) covering dual-mode execution equivalence, two
    whole-system fully-sharded comparisons with 1{,}000-step loss
    alignment, a 505B-parameter MoE scale demonstration, and distributed
    Muon comparisons against three baseline integrations with expert
    coverage and baseline-issue disclosures
    (\S\ref{sect:eval}).
\end{itemize}

\section{Background and Motivation}
\label{sect:background}

\subsection{Supernode Architecture and Two-Tier Networks}
\label{sec:supernode}

Modern NPU supernodes connect hundreds of accelerator dies through a
high-bandwidth intra-supernode fabric and an inter-supernode datacenter network, which features an order of magnitude higher latency and lower bandwidth~\cite{cloudmatrix384}.
This two-tier hierarchy is the hardware premise of every design decision in this
paper: within a supernode, single-parameter messages are already efficient
and latency is low; across supernodes, message count and fusion dominate.
In this paper, we refer to a \emph{machine} (equivalently, a node) as a physical host,
and multiple machines group as a supernode. We refer to \emph{high-bandwidth domain} (fast tier)
as all intra-machine links, and everything leaving the supernode as the slow tier.
% the intra-supernode fabric---including all intra-machine
% links---is the \emph{high-bandwidth domain} (fast tier), and everything
% leaving the supernode is the slow tier.

\subsection{Fully Sharded Data Parallel and HSDP}
\label{sec:fsdp-bg}

\begin{figure}[t]
\centering
\begin{tikzpicture}[x=1cm,y=1cm]
  % --- supernode 0: four dies ---
  \foreach \i in {0,1,2,3}
    \node[fbox, minimum width=0.78cm, minimum height=0.62cm, font=\scriptsize]
      (d0\i) at (-1.35+0.9*\i, 0) {r\i};
  \begin{scope}[on background layer]
  \node[draw=hpgreen!80, rounded corners=2pt, fill=hpgreen!4,
        fit=(d00)(d03), inner sep=7pt,
        label={[hpgreen, font=\scriptsize\bfseries]above:Supernode 0}] (sn0) {};
  \end{scope}
  % --- supernode 1: four dies ---
  \foreach \i [evaluate=\i as \r using int(4+\i)] in {0,1,2,3}
    \node[fbox, minimum width=0.78cm, minimum height=0.62cm, font=\scriptsize]
      (d1\i) at (-1.35+0.9*\i, -2.5) {r\r};
  \begin{scope}[on background layer]
  \node[draw=hpgreen!80, rounded corners=2pt, fill=hpgreen!4,
        fit=(d10)(d13), inner sep=7pt,
        label={[hpgreen, font=\scriptsize\bfseries]below:Supernode 1}] (sn1) {};
  \end{scope}
  % --- shard dimension bracket inside supernode 0 ---
  \draw[gflow, <->] ($(d00.south west)+(-0.04,-0.18)$) --
    node[below, fnote] {Shard dim (intra-supernode): AG / RS}
    ($(d03.south east)+(0.04,-0.18)$);
  % --- inter-supernode (replica) link ---
  \draw[oflow, <->] ($(sn0.south)+(1.15,-0.62)$) --
    ($(sn1.north)+(1.15,0.08)$);
  \node[fnote, text width=3.6cm, anchor=east] at ($(sn0.south)+(-0.1,-0.68)$)
    {Replica dim (inter-supernode, $\sim\!10\times$ slower): fused gradient
     all-reduce};
\end{tikzpicture}
\caption{Two-tier supernode network and the hybrid-sharding mapping. The
shard dimension lives inside a supernode's high-bandwidth fabric; the
replica dimension crosses the slower datacenter tier. Every communication
policy in this paper is derived from this asymmetry. (AG/RS:
all-gather/reduce-scatter.)}
\Description{Diagram of two supernodes of four dies each; the shard dimension with all-gather and reduce-scatter sits inside a supernode, the replica dimension with one fused gradient all-reduce crosses the slower inter-supernode tier.}
\label{fig:topo}
\end{figure}
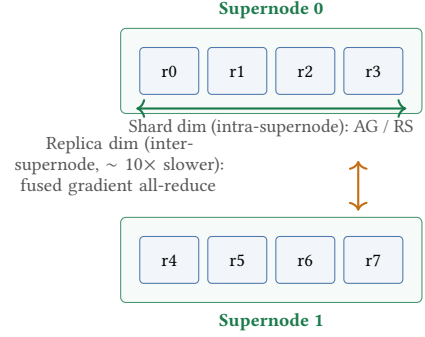

Fully sharded data parallelism (FSDP) shards parameters, gradients, and
optimizer states across the data-parallel group, all-gathers each
parameter on demand, releases it after use, and reduce-scatters gradients
in the backward pass~\cite{zero,pytorchfsdp, disdp_isca26}. Hybrid sharded data
parallelism (HSDP) adds
a replica dimension: parameters are sharded within each group and
replicated across groups, with a cross-replica gradient all-reduce
following the intra-group reduce-scatter~\cite{hsdp}. The canonical mapping
onto supernode hardware places the sharding dimension inside a supernode
and the replica dimension across supernodes, matching communication cost to
link speed~\cite{megascale} (Figure~\ref{fig:topo}).

\subsection{The Muon Optimizer and Newton--Schulz Iteration}
\label{sec:bg-muon}

Muon updates a weight matrix $W \in \mathbb{R}^{m \times n}$ by
accumulating momentum $G$, orthogonalizing the update direction with a
Newton--Schulz (NS) iteration
\begin{equation}
\begin{aligned}
X_{k+1} &= a_k X_k + b_k\,(X_k X_k^{\top}) X_k + c_k\,(X_k X_k^{\top})^2 X_k,\\
X_0 &= G / (\lVert G \rVert_F + \epsilon),
\end{aligned}
\end{equation}
which approximates $\mathrm{msign}(G) = UV^{\top}$ from the SVD, and
applying an RMS-matched scaling $0.2\sqrt{\max(m, n)}$ so that the AdamW
learning-rate scale transfers directly~\cite{muon,moonlight}. The per-step
coefficients $(a_k, b_k, c_k)$ define the variant: \sys defaults to an
asymmetric five-step variant and retains the legacy constant-coefficient
quintic as a compatibility option (\S\ref{sec:batched-ns}). Two properties are
decisive for distribution. First, Muon applies only to matrices:
two-dimensional parameters directly, higher-order parameters sliced along
their matrix planes, and one-dimensional parameters not at all (they are
delegated to AdamW). Second, orthogonalization is \emph{global}: the Gram factor
couples one matrix dimension in its entirety, so a shard produced by any
row- or column-wise cut cannot in general be orthogonalized in isolation. Muon has been validated at frontier
scale, including Kimi K2's MuonClip variant~\cite{kimik2}, Kimi
K3~\cite{kimik3}, DeepSeek-V4 up to the 1.6T V4-Pro~\cite{deepseekv4}, and
the GLM-4.5/5 series~\cite{glm45}, and is now supported by DeepSpeed and
Megatron-LM~\cite{megatronlm}; it has become an increasingly adopted
alternative to
AdamW for frontier pretraining.

\subsection{Distributed Tensors and Autodiff}
\label{sec:dtensor-bg}

PyTorch's native distributed tensor performs operator-level distributed
dispatch \emph{below} autograd~\cite{pytorch2,pytorchdtensor}: every
operator invocation passes through dispatch and distributed-metadata
handling, while sharding propagation may hit cached Python/C++ fast paths
and redistribution is inserted only when the propagated layout requires
it. %---the exact split is version-dependent. 
This generality comes at two costs: a per-operator
dispatch and metadata overhead paid on every training step even on the warm
path, and a single execution
mode that provides no validate/production dual-mode contract sharing one
static plan. Declarative SPMD frameworks such as TorchTitan and
NeMo-AutoModel build on this primitive~\cite{torchtitan,nemoautomodel}:
they achieve the decoupling of parallelization from model code that
Megatron-style frameworks lack~\cite{megatronlm}, but inherit the
dispatch overhead everywhere.

\subsection{Motivation: Design Pain Points}
\label{sec:motivation}

This paper is motivated by the three main problems of current PyTorch: 
First, the packing and unpacking passes around the fused collectives of the
evaluated PyTorch 2.9 FSDP2 configuration burn
memory bandwidth linearly in the number of layers; in
communication-dominated configurations, host-side fusion alone may provide
limited end-to-end benefit (\S\ref{sec:zerocopy}). Second, naive distributed Muon
repeats the orthogonalization once per shard rank, and again per replica
under hybrid sharding; moreover, small matrices executed one by one are
host-bound in dispatch rather than arithmetic
(\S\ref{sec:muon-challenge}). Third, waiting on
inter-supernode reductions inside the backward pass forces faster
intra-supernode collectives to queue behind the slowest traffic
(\S\ref{sec:coexist}).

\section{System Overview}
\label{sec:overview}

\begin{figure}[t]
\centering
\begin{tikzpicture}[x=1cm,y=1cm]
  \node[ghost, text width=7.3cm] (in) at (0,0)
    {\textbf{Model Structure + Device Mesh + Declarative User Overrides}};
  \node[fbox=hporange, text width=7.3cm] (plan) at (0,-1.5)
    {\textbf{One Static Sharding Plan} (\S\ref{sec:planner})\hfill
     {\tiny\bfseries\textcolor{hpgray}{[PyTorch]}}\\[1pt]
     {\scriptsize auto-derived, override-combined, statically linted:
      per-parameter placements, boundary transitions, layout contracts;
      one plan drives both production and validation modes}};
  \node[fbox, text width=2.26cm, minimum height=104pt] (dt) at (-2.52,-4.4)
    {\textbf{Dual-Mode Distributed-Tensor Execution} (\S\ref{sect:dtensor})\\[1pt]
     {\scriptsize boundary contracts; production / validation modes}\\[2pt]
     {\tiny\bfseries\textcolor{hpgray}{[PyTorch]}}};
  \node[good, text width=2.26cm, minimum height=104pt] (fsdp) at (0,-4.4)
    {\textbf{Topology-Aware FSDP/HSDP} (\S\ref{sect:fsdp})\\[1pt]
     {\scriptsize two-tier collectives from the mesh-to-topology mapping;
      backward pipeline}\\[2pt]
     {\tiny\bfseries\textcolor{hpgray}{[PyTorch / MindSpore]}}};
  \node[good, text width=2.26cm, minimum height=104pt] (opt) at (2.52,-4.4)
    {\textbf{Layout-Driven Distributed Muon} (\S\ref{sect:muon})\\[1pt]
     {\scriptsize grouping, owners, and dedup domains from layouts;
      Muon + AdamW}\\[2pt]
     {\tiny\bfseries\textcolor{hpgray}{[PyTorch]}}};
  \node[ghost, text width=7.3cm] (plat) at (0,-7.15)
    {\textbf{Shared Platform Abstraction + Collective Runtime}:
     PyTorch / MindSpore backends; vendor collective-communication
     library};
  \draw[flow] (in) -- (plan);
  \draw[flow] (plan.south -| dt.north) --
    node[left, flabel] {contracts} (dt.north);
  \draw[flow] (plan.south) --
    node[right, flabel] {topology} (fsdp.north);
  \draw[flow] (plan.south -| opt.north) --
    node[right, flabel] {layouts} (opt.north);
  \draw[flow] (dt.south) -- (dt.south |- plat.north);
  \draw[flow] (fsdp.south) -- (fsdp.south |- plat.north);
  \draw[flow] (opt.south) -- (opt.south |- plat.north);
\end{tikzpicture}
\caption{Architecture of \sys: First is one static sharding plan derived from the
model structure, the device mesh, and declarative user
overrides. The plan materializes the placements and contracts consumed, directly
or indirectly, by three subsystems: (1)~Dual-mode distributed-tensor execution
(boundary contracts; production and validation modes share the same plan),
(2)~topology-aware FSDP/HSDP (the two-tier mapping shapes its collective
domains), and (3)~layout-driven distributed Muon (layouts shape its grouping
and dedup domains). Backend badges mark where each component runs: the
core distributed-tensor and fully-sharded layers run on both backends,
while the planner, the above-autograd interception, and the distributed
optimizer are implemented on the PyTorch backend and evaluated in this
paper. The platform abstraction and collective runtime at the bottom are
shared infrastructure, not a pipeline stage; symmetric memory (not shown)
serves only specific fused TP/EP kernels. (TP/EP: tensor/expert
parallelism.)}
\label{fig:arch}
\end{figure}
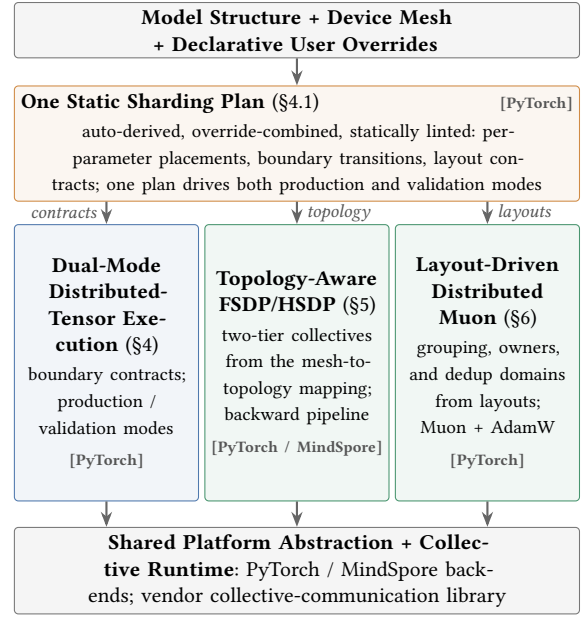

\sys is organized around a single static artifact
(Figure~\ref{fig:arch}): a declarative \emph{sharding plan} derived from
the model structure
(\S\ref{sec:planner}) and combined with declarative user overrides. The plan materializes
the placements and boundary contracts consumed by three subsystems either directly or indirectly. The
\emph{dual-mode distributed-tensor runtime} (\S\ref{sec:dualmode}) executes
its boundary transitions: in
production mode parameters are permanently materialized as plain local
tensors and only precompiled boundary collectives execute; in validation mode
layout metadata propagates end-to-end with fail-fast contract checking.
The \emph{topology-aware fully-sharded layer} (\S\ref{sect:fsdp})
reads the plan's distribution semantics to run per-parameter zero-copy
collectives within a supernode and fused
reduction across supernodes. The \emph{distributed optimizer
layer} (\S\ref{sect:muon}) derives grouping, compute owners, and dedup
domains from the plan's placements, running layout-driven Muon for matrix
parameters chained with AdamW for the rest.

The three subsystems share a platform abstraction and a collective
runtime. The platform abstraction spans the PyTorch and MindSpore
backends. The core distributed-tensor and fully-sharded layers run on
both, while the dual-mode planner, the above-autograd interception
(\S\ref{sec:autodiff}), and distributed Muon are implemented on the
PyTorch backend; this paper evaluates the PyTorch 2.9/Ascend path, and the
MindSpore backend is an engineering capability rather than an equally
evaluated target of the three contributions. The collective runtime is the
vendor collective-communication library. Separately, a symmetric-memory
facility serves only specific fused tensor- and expert-parallel kernels;
it is not a shared dependency of the three designs.

Three design principles recur throughout. \emph{Model/system decoupling}:
the strategy is declarative data, injected without touching model code.
\emph{Explicit, verifiable semantics}: every distribution fact lives in the
plan and is checkable in validation mode. \emph{Topology awareness
everywhere}: communication policy, dedup domains, and pipelining are all
derived from the two-tier network structure rather than tuned per model.
Tables~\ref{tab:feature} and~\ref{tab:muonfeat} position \sys against the
PyTorch-native stack and prior distributed Muon along these principles.

\begin{table}[t]
\caption{Distribution semantics and fully-sharded communication: \sys
versus the evaluated PyTorch DTensor + FSDP2 configuration (version to be
pinned in \S\ref{sec:setup}). ``--'' marks a feature not provided by that
baseline configuration.}
\label{tab:feature}
{\footnotesize
\setlength{\tabcolsep}{3pt}
\begin{tabular}{@{}>{\RaggedRight\arraybackslash}p{3.6cm}%
>{\centering\arraybackslash}p{1.9cm}%
>{\centering\arraybackslash}p{1.6cm}@{}}
\toprule
 & Evaluated PyTorch DTensor + FSDP2 config.\ & \sys \\
\midrule
\multicolumn{3}{@{}l}{\emph{Distribution semantics}} \\
Declarative sharding plan & manual & auto-derived \\
Zero steady-state per-operator DTensor dispatch & -- & \checkmark \\
Fail-fast validation mode & -- & \checkmark \\
\addlinespace
\multicolumn{3}{@{}l}{\emph{Fully-sharded communication}} \\
Zero-copy per-parameter collectives (leading-dim layouts) & -- & \checkmark \\
Two-tier (supernode-aware) policy & manual mesh & \checkmark \\
No layer-local wait on slow tier & -- & \checkmark \\
\bottomrule
\end{tabular}}
\end{table}

\begin{table}[t]
\caption{Distributed Muon: \sys versus Moonlight's distributed
Muon~\cite{moonlight} (bucket-based, ZeRO-1-style). ``--'' marks a feature not provided by the
baseline.}
\label{tab:muonfeat}
{\footnotesize
\setlength{\tabcolsep}{3pt}
\begin{tabular}{@{}>{\RaggedRight\arraybackslash}p{3.6cm}%
>{\centering\arraybackslash}p{1.9cm}%
>{\centering\arraybackslash}p{1.6cm}@{}}
\toprule
 & Moonlight distr.\ Muon & \sys \\
\midrule
Comm.\ domains from layout semantics & -- & \checkmark \\
Shard + replica two-level dedup & shard-only & \checkmark \\
Shape-fused batched orthogonalization & -- & \checkmark \\
\bottomrule
\end{tabular}}
\end{table}

\section{Dual-Mode Distributed-Tensor Execution and Autodiff}
\label{sect:dtensor}

This section presents how \sys executes and validates a sharding plan. We
first describe how the plan is derived automatically from the model
structure (\S\ref{sec:planner}) and how boundary communication is
precompiled when the plan is applied (\S\ref{sec:boundary}). We then
present the production and validation execution modes
(\S\ref{sec:dualmode}), explain the architectural decision that makes them
possible by intercepting \emph{above} autograd rather than below it
(\S\ref{sec:autodiff}), and close with the numerical harness that turns
the two modes into a verification loop (\S\ref{sec:gradequiv}).

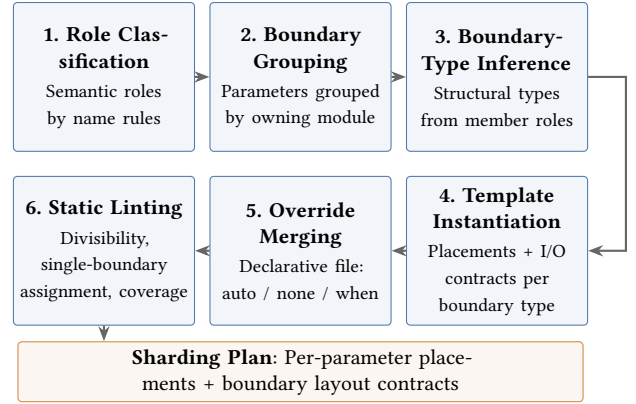
\begin{figure}[t]
\centering
\begin{tikzpicture}[x=1cm,y=1cm]
  \node[fbox, text width=2.22cm, minimum height=56pt] (s1) at (-2.6,0)
    {\textbf{1. Role Classification}\\[1pt]
     {\scriptsize Semantic roles by name rules}};
  \node[fbox, text width=2.22cm, minimum height=56pt] (s2) at (0,0)
    {\textbf{2. Boundary Grouping}\\[1pt]
     {\scriptsize Parameters grouped by owning module}};
  \node[fbox, text width=2.22cm, minimum height=56pt] (s3) at (2.6,0)
    {\textbf{3. Boundary-Type Inference}\\[1pt]
     {\scriptsize Structural types from member roles}};
  \node[fbox, text width=2.22cm, minimum height=56pt] (s4) at (2.6,-2.3)
    {\textbf{4. Template Instantiation}\\[1pt]
     {\scriptsize Placements + I/O contracts per boundary type}};
  \node[fbox, text width=2.22cm, minimum height=56pt] (s5) at (0,-2.3)
    {\textbf{5. Override Merging}\\[1pt]
     {\scriptsize Declarative file: auto / none / when}};
  \node[fbox, text width=2.22cm, minimum height=56pt] (s6) at (-2.6,-2.3)
    {\textbf{6. Static Linting}\\[1pt]
     {\scriptsize Divisibility, single-boundary assignment, coverage}};
  \node[fbox=hporange, text width=7.3cm] (out) at (0,-3.9)
    {\textbf{Sharding Plan}: Per-parameter placements + boundary layout
     contracts};
  \draw[flow] (s1) -- (s2);
  \draw[flow] (s2) -- (s3);
  % serpentine turns routed outside the box grid
  \draw[flow] (s3.east) -- ++(0.5,0) |- (s4.east);
  \draw[flow] (s4) -- (s5);
  \draw[flow] (s5) -- (s6);
  \draw[flow] (s6.south) -- (s6.south |- out.north);
\end{tikzpicture}
\caption{The six logical passes of the sharding planner. The plan is a
static, data-level artifact: machine-checkable before a single collective
is launched.}
\Description{Flow diagram of the six logical planner passes (role classification, boundary grouping, boundary-type inference, template instantiation, override merge, static lint) producing a sharding plan of placements and boundary contracts; the pass-three-to-four turn routes outside the box grid.}
\label{fig:planner}
\end{figure}

\subsection{Declarative Sharding Planning}
\label{sec:planner}

Manually authored, model-specific parallelization plans scale poorly with
model diversity: each
new architecture requires its own annotation of which parameter is sharded
along which device-mesh dimension, and the annotations are intertwined with
model code. We observe, however, that the sharding of a supported model
family is determined almost entirely by the \emph{semantic role} each
parameter plays (e.g., column-parallel projection, normalization scale, and
expert weight), rather than by architecture-specific detail. \sys therefore
derives the sharding plan statically from the module graph.

A sharding plan assigns to every parameter a placement over the device mesh
and to every module boundary a pair of input/output layout contracts. The
planner produces this plan in six logical passes
(Figure~\ref{fig:planner}). (1)~\emph{Role Classification}:
each parameter is classified into a fixed semantic-role taxonomy (e.g.,
column-wise or row-wise projection, normalization, embedding, routed or
shared expert, fused QKV) by segment-aware matching of its qualified name
against a small rule set, optionally refined by per-architecture overrides.
(2)~\emph{Boundary Grouping}: parameters are grouped by their owning
module, and each group forms a candidate communication boundary.
(3)~\emph{Boundary-Type Inference}: each boundary is classified into a
structural type (attention, MLP, normalization, embedding, output head,
MoE routing, MoE expert) from the roles of its members.
(4)~\emph{Template Instantiation}: boundary templates, one per
boundary type, map roles to concrete placements and generate the
input/output layout contracts for every active mesh dimension.
(5)~\emph{Override Merging}: a declarative override file, with the
sentinels \emph{auto} (derive automatically), \emph{none} (leave local),
and \emph{when} (conditional clause), is merged over the derived plan, so
users state only what deviates from the default.
(6)~\emph{Static Linting}: the merged plan is checked for divisibility of
sharded dimensions, assignment of every parameter to exactly one boundary
specification, and complete coverage
of trainable parameters; any violation aborts before a single collective is
launched. The role taxonomy and boundary templates cover the supported
Transformer families (dense attention/MLP, MoE, and MLA-style
factorized projections); parameters that match no rule fall back to the
override file, so coverage degrades gracefully to manual specification
rather than failing silently.

The effect is that the repeated per-model parallelization rules of
manually authored plans are centralized into a small set of naming rules. 
Because the plan is a static, data-level artifact, it becomes
machine-checkable before execution.

\subsection{Precompiled Boundary Communication}
\label{sec:boundary}

The layout transformations between adjacent modules are the only
communication the sharding plan's forward contracts induce in
production-mode execution. \sys materializes every boundary's
transformation into a precompiled \emph{transition} at plan-application time: 
Transitions in the supported set are lowered to concrete
collective sequences (tensor-parallel collectives further to
differentiable local-tensor collective primitives), which are cached and replayed
unchanged at every step; all other transitions retain a generic
distributed-tensor redistribution fallback, so expressiveness is never
gated by the lowerer.

This design occupies a deliberate middle point between two extremes. Eager
per-operator dispatch, as in PyTorch's native distributed tensor, retains
per-operator distributed dispatch and metadata handling at every operator
invocation; whole-graph compilation, as in
XLA-style auto-parallel systems, removes dispatch cost but forfeits eager
debuggability and model flexibility. \sys keeps eager execution \emph{within}
modules. In validation mode, an operator still passes through interception,
argument preprocessing, cache-keyed layout inference, local execution, and
output checking, with the cache eliminating only repeated layout inference,
while production mode runs none of this.
Meanwhile, \sys precompiles communication \emph{across} module boundaries, 
which is where all production traffic induced by the plan's
forward layout contracts occurs.

Two scoping remarks complete the picture. First, a fallback transition in
production mode executes as generic redistribution directly on plain local
tensors through the same collective library, so it never re-enters the
wrapper machinery; only validation mode routes it through the generic
distributed-tensor path so that layouts remain tracked. Second,
communication inside user-declared opaque regions, such as expert all-to-all
routing, context-parallel attention, and custom kernels, belongs to those
regions' own declared contracts, not to the planner's forward-contract
traffic accounted for above.

\subsection{Production and Validation Modes}
\label{sec:dualmode}

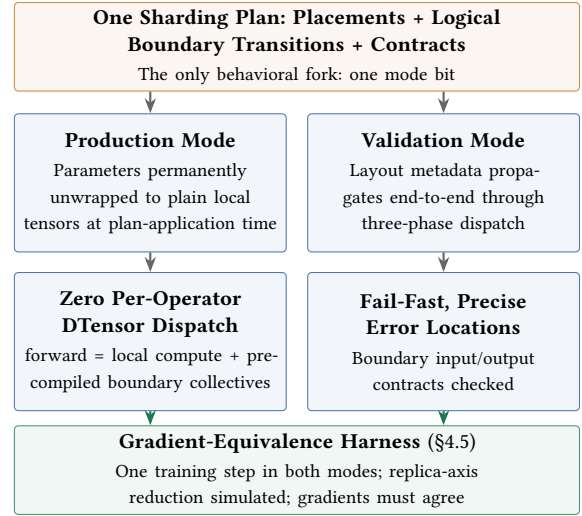
\begin{figure}[t]
\centering
\begin{tikzpicture}[x=1cm,y=1cm]
  \node[fbox=hporange, text width=7.3cm] (plan) at (0,0)
    {\textbf{One Sharding Plan: Placements + \mbox{Logical} Boundary
     Transitions + Contracts}\\[1pt]
     {\scriptsize The only behavioral fork: one mode bit}};
  \node[fbox, text width=3.42cm, minimum height=52pt] (p1) at (-1.94,-1.8)
    {\textbf{Production Mode}\\[1pt]
     {\scriptsize Parameters permanently unwrapped to plain local tensors
      at plan-application time}};
  \node[fbox, text width=3.42cm, minimum height=52pt] (p2) at (-1.94,-3.9)
    {\textbf{Zero Per-Operator DTensor Dispatch}\\[1pt]
     {\scriptsize forward = local compute + precompiled boundary
      collectives}};
  \node[fbox, text width=3.42cm, minimum height=52pt] (v1) at (1.94,-1.8)
    {\textbf{Validation Mode}\\[1pt]
     {\scriptsize Layout metadata propagates end-to-end through three-phase
      dispatch}};
  \node[fbox, text width=3.42cm, minimum height=52pt] (v2) at (1.94,-3.9)
    {\textbf{Fail-Fast, Precise Error Locations}\\[1pt]
     {\scriptsize Boundary input/output contracts checked}};
  \node[good, text width=7.3cm] (h) at (0,-5.6)
    {\textbf{Gradient-Equivalence Harness} (\S\ref{sec:gradequiv})\\[1pt]
     {\scriptsize One training step in both modes; replica-axis reduction
      simulated; gradients must agree}};
  \draw[flow] (plan.south -| p1.north) -- (p1.north);
  \draw[flow] (plan.south -| v1.north) -- (v1.north);
  \draw[flow] (p1) -- (p2);
  \draw[flow] (v1) -- (v2);
  \draw[gflow, dashed] (p2.south) -- (p2.south |- h.north);
  \draw[gflow, dashed] (v2.south) -- (v2.south |- h.north);
\end{tikzpicture}
\caption{Dual-mode execution. Both modes share the same plan, the same
logical boundary transitions, and the same contracts (concrete primitives
may differ at fallback transitions), so a validate-mode pass is evidence
about what production executes; the gradient-equivalence harness closes the
loop at the value level.}
\Description{Diagram of one sharding plan forking by a single mode bit into production mode (plain local tensors, zero per-operator DTensor dispatch) and validation mode (end-to-end layout propagation, fail-fast checks), both feeding a gradient-equivalence harness.}
\label{fig:dualmode}
\end{figure}

A single sharding plan, with its logical boundary transitions and
contracts, drives two execution modes (Figure~\ref{fig:dualmode}): a
\emph{production mode} and a \emph{validation mode} (code mode name
\texttt{validate}); the only
behavioral fork is one mode bit in the execution plan. The two modes share
the plan, the logical transitions, and the contracts; the concrete
primitives may differ where a transition takes the generic redistribution
fallback (production can use the local lowerer where validate redistributes
through the generic distributed-tensor path).

In \emph{production mode}, parameters are permanently materialized as plain
local tensors when the plan is applied. The forward pass performs purely
local computation stitched together by the precompiled boundary collectives;
no layout inference, metadata propagation, or wrapper dispatch executes at
all, so production mode pays no steady-state per-operator DTensor dispatch.

In \emph{validation mode}, tensors carry their layout metadata end-to-end.
Every operator invocation propagates layouts through the distributed-tensor
semantics, and every module boundary checks its input and output contracts;
a mismatch raises an immediate, precisely located error rather than a
downstream numerical corruption.

Because both modes execute the \emph{same} plan with the \emph{same}
logical boundary transitions and contracts, a validate-mode pass over a
model is strong evidence
that production mode executes the intended sharding: verification is a
property of the plan, not of a separate code path. User-injected
regions, including expert all-to-all routing, context-parallel attention, 
and custom kernels, must declare themselves under a \emph{region axiom}: Either
\emph{transparent}, in which case their outputs are checked against the
downstream contract; or \emph{opaque}, in which case they are treated as
entrusted black boxes whose layout obligations are explicit in the plan.

\subsection{API-Level Sharding Semantics: Above Autograd, Not Below}
\label{sec:autodiff}

{\sloppy
The dual-mode design is enabled by one architectural decision: \sys
intercepts at the tensor API layer \emph{above} autograd by a function-level
callback that fires before the autograd engine records the operation
(PyTorch's \texttt{\_\_torch\_function\_\_}), whereas PyTorch's native
distributed tensor dispatches at the operator layer \emph{below}
autograd, inside the recorded graph (the exact Python/C++ dispatch path is
version-dependent~\cite{pytorchdtensor}).
Four consequences follow.
\par}

First, the distributed tensor describes \emph{only} forward sharding
semantics. It is an auxiliary, API-level description of how data is laid
out, rather than the model's core data structure.

Second, graph construction, differentiation, and kernel execution always
observe plain local tensors. The wrapper never enters the operator layer,
so the autograd engine and the device kernels require no awareness of
distribution whatsoever.

Third, switching between production and validation modes is a metadata-level
operation: because no operator-level machinery depends on the wrapper,
production mode can discard it entirely at plan-application time.

Fourth, distribution semantics and differentiation semantics are orthogonal
by construction. Forward layout inference handles distribution by
a three-phase dispatch that preprocesses arguments, 
infers the output layout through a cached rule
table, and expands to local execution; ordinary
local autograd handles differentiation. No distributed autograd engine is
needed. Gradients materialize on local shards, where the data-parallel
hooks reduce-scatter them (\S\ref{sect:fsdp}); residual gradients along
tensor-parallel replica axes are reduced exactly once at the end of the
backward pass, preserving the mathematical semantics of a replicated
computation.

The placement algebra itself is correspondingly richer than the
shard/replicate/partial trio of the native design: it supports strided and
ragged sharding, uneven (ceil-chunk) sharding for dimensions not divisible
by the mesh size, and shard-aware random number generation with per-shard
offset tracking so that stochastic layers remain bit-reproducible under
resharding.

\subsection{Gradient Equivalence Validation}
\label{sec:gradequiv}

Layout checking alone does not establish that production mode computes the
\emph{same values} as the semantics describe. \sys therefore ships a
numerical harness that lifts validation mode from layout checking to value
checking: it runs one training step in both modes, simulates within
validation mode the replica-axis gradient reduction that production performs
inside the data-parallel hooks, and asserts that the resulting gradients
agree. This closes the loop: \emph{Declared} (the plan) $=$
\emph{verified} (validation mode) $=$ \emph{what production runs}. The
harness is a design and engineering verification mechanism over the
supported model architectures, not a complete numerical-equivalence proof;
\S\ref{sec:eval-dtensor} presents the end-to-end evidence: over the
evaluated 60-step horizon the two modes execute identical device work with
identical loss and gradient-norm trajectories, and \S\ref{sec:eval-fsdp}
adds a 1{,}000-step whole-system run whose per-step losses track the
baseline at Pearson $r>0.999997$.
   % dual-mode DTensor + distributed autodiff

\section{Topology-Aware Fully-Sharded Parallelism}
\label{sect:fsdp}

This section describes how \sys materializes sharded parameters and
reduces gradients over a two-tier HSDP mesh. The design premise is the
two-tier network of supernode hardware (\S\ref{sec:supernode}), a two-tier
communication hierarchy in which, within a
supernode, links are fast enough that single-parameter messages are already
efficient; across supernodes, bandwidth is an order of magnitude scarcer
and message count dominates. Every choice below follows from treating the
two tiers differently: per-parameter zero-copy collectives within a
supernode (\S\ref{sec:zerocopy}), fused reduction across supernodes
(\S\ref{sec:twotier}), a backward schedule whose layer-level path never
waits on the slow tier (\S\ref{sec:pipeline}, \S\ref{sec:coexist}), and the
engineering capabilities that make the scheme practical
(\S\ref{sec:fsdp-eng}).

\begin{figure*}[t]
\centering
\begin{tikzpicture}[x=1cm,y=1cm]
  % ---------- (a) fused path ----------
  \node[font=\footnotesize\bfseries, text=hpred, anchor=west]
    at (-8.3,1.15) {(a) Fused Path (evaluated PyTorch 2.9 FSDP2
    configuration): packing and view reconstruction around an all-gather};
  \node[ghost, text width=1.7cm, minimum height=58pt] (shards) at (-7.9,-0.35)
    {\scriptsize $p_1$ shard\\[2pt] $p_2$ shard\\[2pt] $p_3$ shard\\[2pt]
     $p_4$ shard};
  \node[bad, text width=1.35cm, minimum height=58pt] (copyin) at (-5.75,-0.35)
    {\scriptsize Copyin (pack)};
  \node[ghost, text width=1.55cm, minimum height=58pt] (fin) at (-3.8,-0.35)
    {\scriptsize Fused input buffer};
  \node[fbox, minimum width=0.85cm, minimum height=58pt] (ag) at (-2.15,-0.35)
    {\scriptsize AG};
  \node[ghost, text width=1.55cm, minimum height=58pt] (fout) at (-0.4,-0.35)
    {\scriptsize Fused output buffer};
  \node[bad, text width=1.35cm, minimum height=58pt] (copyout) at (1.75,-0.35)
    {\scriptsize Copyout (unpack)};
  \node[ghost, text width=1.9cm, minimum height=58pt] (full) at (3.95,-0.35)
    {\scriptsize $p_1$..$p_4$ full-parameter buffers};
  \draw[flow] (shards) -- (copyin);
  \draw[flow] (copyin) -- (fin);
  \draw[flow] (fin) -- (ag);
  \draw[flow] (ag) -- (fout);
  \draw[flow] (fout) -- (copyout);
  \draw[flow] (copyout) -- (full);
  \node[fnote, anchor=west, text width=3.1cm] at (5.3,-0.35)
    {\scriptsize Copy kernels occupy the compute stream};
  % ---------- (b) per-parameter zero-copy ----------
  \node[font=\footnotesize\bfseries, text=hpgreen, anchor=west]
    at (-8.3,-2.0) {(b) Per-Parameter Zero-Copy Path (\sys): no data
    movement around the collective};
  \node[good, text width=1.7cm, minimum height=34pt] (pshard) at (-7.3,-3.1)
    {\scriptsize $p_i$ shard\\ \scriptsize(the parameter's\\ \scriptsize own storage)};
  \node[fbox, minimum width=0.85cm, minimum height=34pt] (ag2) at (-4.6,-3.1)
    {\scriptsize AG};
  \node[good, text width=2.6cm, minimum height=34pt] (pbuf) at (-1.6,-3.1)
    {\scriptsize Persistent per-parameter buffer;\\ \scriptsize full $p_i$ is
     a \emph{view} of it};
  \draw[gflow] (pshard) -- (ag2);
  \draw[gflow] (ag2) -- (pbuf);
  \node[fnote, anchor=west, text width=6.2cm] at (0.4,-3.1)
    {\scriptsize Input \emph{is} the shard itself: zero copyin; output lands
     in the parameter's own buffer: zero copyout; collective payload volume
     identical to (a)};
\end{tikzpicture}
\caption{All-gather data paths within a supernode, for the evaluated
PyTorch 2.9 FSDP2 configuration and \sys. (a)~The fused scheme of
the evaluated baseline
surrounds the all-gather with packing and view-reconstruction passes over
device memory. (b)~\sys issues one collective per parameter and removes both
copies for leading-dimension-sharded layouts; the full parameter is a view
of its persistent buffer.}
\Description{Two data-path diagrams for all-gather: the fused baseline path with copy-in and copy-out passes around a fused buffer, versus the per-parameter zero-copy path where the shard is the collective input and the full parameter is a view of its buffer.}
\label{fig:zerocopy}
\end{figure*}

\begin{figure*}[t]
\centering
\begin{tikzpicture}[x=1cm,y=1cm]
  % ---------- (a) two-tier policy ----------
  \node[font=\footnotesize\bfseries, text=hpblue, anchor=north west,
        text width=9.2cm, align=left]
    at (-8.6,1.45) {(a) Split policy: many small zero-copy collectives
    inside, one fused collective across};
  \foreach \sn/\xs/\rb in {0/-8.0/0, 1/-3.3/4} {
    \foreach \i [evaluate=\i as \r using int(\rb+\i)] in {0,1,2,3}
      \node[fbox, minimum width=0.62cm, minimum height=0.5cm,
            font=\scriptsize] (r\sn\i) at (\xs+0.75*\i, -0.55) {r\r};
    \begin{scope}[on background layer]
    \node[draw=hpgreen!80, rounded corners=2pt, fill=hpgreen!4,
          fit=(r\sn0)(r\sn3), inner sep=6pt,
          label={[hpgreen, font=\scriptsize\bfseries]above:Supernode \sn}] (snb\sn) {};
    \end{scope}
  }
  \draw[gflow, <->] ($(r00.south west)+(0,-0.32)$) -- ($(r03.south east)+(0,-0.32)$);
  \node[fnote] at (-6.9,-1.6) {Intra: per-param AG/RS,\\ zero-copy};
  \draw[gflow, <->] ($(r10.south west)+(0,-0.32)$) -- ($(r13.south east)+(0,-0.32)$);
  \node[fnote] at (-2.2,-1.6) {Intra: per-param AG/RS,\\ zero-copy};
  \draw[oflow, <->] ($(snb0.east)+(0.08,0)$) -- ($(snb1.west)+(-0.08,0)$);
  \node[fnote, fill=white, inner sep=1.5pt, align=center] at (-4.65,0.55)
    {Inter: one fused all-reduce (async)};
  % ---------- (b) fused buffer mechanics ----------
  \node[font=\footnotesize\bfseries, text=hpblue, anchor=north west,
        text width=7.6cm, align=left]
    at (1.0,1.45) {(b) Fusion without copies: reduce-scatter writes buffer
    views};
  \node[fbox, text width=1.15cm, minimum height=30pt] (rs1) at (2.15,0.1)
    {\scriptsize RS $p_1$};
  \node[fbox, text width=1.15cm, minimum height=30pt] (rs2) at (3.75,0.1)
    {\scriptsize RS $p_2$};
  \node[fbox, text width=1.15cm, minimum height=30pt] (rs3) at (5.35,0.1)
    {\scriptsize RS $p_3$};
  \node[good, text width=1.15cm, minimum height=30pt] (v1) at (2.15,-1.25)
    {\scriptsize $p_1$ view};
  \node[good, text width=1.15cm, minimum height=30pt] (v2) at (3.75,-1.25)
    {\scriptsize $p_2$ view};
  \node[good, text width=1.15cm, minimum height=30pt] (v3) at (5.35,-1.25)
    {\scriptsize $p_3$ view};
  \node[ghost, text width=0.75cm, minimum height=30pt] (pad) at (6.6,-1.25)
    {\scriptsize pad};
  \draw[flow] (rs1) -- (v1);
  \draw[flow] (rs2) -- (v2);
  \draw[flow] (rs3) -- (v3);
  \draw[oflow, <->] ($(v1.south west)+(0,-0.22)$) --
    node[below, fnote, align=center, text width=6.2cm] {one
    \texttt{all\_reduce} over the whole buffer (SUM), asynchronous}
    ($(pad.south east)+(0,-0.22)$);
  \node[fnote, align=center] at (4.4,-3.15)
    {Buffer base 512B-aligned; total length padded, pre-zeroed:\\
     padding is neutral under SUM; gradients read back from views. \\
     Zero copyin, zero copyout};
\end{tikzpicture}
\caption{Two-tier hybrid sharding. (a)~Per-parameter zero-copy collectives
run within each supernode; a single fused all-reduce crosses supernodes per
unit. (b)~The fusion itself is copy-free: per-parameter reduce-scatter
outputs land directly in their corresponding views of one contiguous
buffer with a 512-byte-aligned base address and a padded total length,
and one collective reduces the whole buffer. (AG/RS: all-gather/reduce-scatter.)}
\Description{Two-part diagram: (a) per-parameter zero-copy collectives inside each supernode and one fused all-reduce across supernodes; (b) reduce-scatter outputs landing in corresponding views of one pre-zeroed contiguous buffer with a 512-byte-aligned base address and a padded total length, reduced by a single all-reduce.}
\label{fig:hsdp}
\end{figure*}

\subsection{Per-Parameter Zero-Copy Communication}
\label{sec:zerocopy}

The evaluated PyTorch 2.9 FSDP2 baseline configuration
(\S\ref{sec:setup})~\cite{pytorchfsdp,pytorchfsdp2} implements
fused-unit communication with explicit packing. For all-gather, a copy-in
pass packs every shard of a unit into one contiguous buffer before the
collective, and a copy-out pass unpacks the result back into per-parameter
storage afterwards; for reduce-scatter, gradient shards are likewise packed
before the collective, and the fused output is then copied back (or
exposed as views) per parameter. The copy-out pass sits on the critical
path because downstream computation depends on it. We state this data path
as configured in our evaluation; without trace evidence we do not claim
that every copy is exposed on the critical path. Fusion amortizes message
launch overhead, which especially matters on general-purpose Ethernet, for
which the design was made. % (FSDP2's exact packing and memory behavior is
% version- and layout-dependent.)

Within a supernode, this trade inverts. Links are high-bandwidth and
low-latency, so a single parameter's shard is already a large, efficient
message: all-gather is prefetched, while gradient reduction is issued as
an asynchronous reduce-scatter, so communication can overlap with
computation. Fusion's launch savings therefore provide little benefit in
this regime while its copy cost is
real. \sys therefore issues collectives \emph{per parameter} and eliminates
the copies for the leading-dimension sharded layouts that dominate in
practice: the all-gather input \emph{is} each parameter's own shard, and
the output is written directly into a persistent per-parameter buffer of
which the materialized full parameter is merely a view; reduce-scatter is
symmetric. (Parameters sharded on other dimensions take a
chunk-and-concatenate copy-out fallback.) The collective payload volume is
unchanged; the path removes the surrounding device-memory packing and
unpacking traffic in the eligible layouts
(Figure~\ref{fig:zerocopy}).

\subsection{Two-Tier HSDP: Fused Reduction without Copies}
\label{sec:twotier}

Across supernodes the trade inverts again: at the inter-supernode
bandwidths and message sizes of the configurations evaluated here, fewer
and larger messages are generally preferable. \sys maps hybrid sharded data parallelism onto the
two tiers with a split policy: \sys performs per-parameter all-gather and reduce-scatter
\emph{within} each supernode (zero-copy), one fused all-reduce \emph{across}
supernodes per unit (Figure~\ref{fig:hsdp}).

Crucially, the fusion itself is zero-copy. A contiguous gradient buffer is
allocated per unit in each step; the allocation's base address is 512-byte
aligned and the total buffer length is padded to a 512-byte multiple.
Padding is zero-initialized and is therefore neutral under SUM. Each
reduce-scatter writes its output directly into the corresponding view of
the fused buffer. One fused all-reduce then covers the whole
buffer, and each parameter reads its final gradient back from its own
view, thus no parameter ever moves. Intra-supernode traffic
trades zero copies for bandwidth, inter-supernode traffic trades fusion for
message count, and the shared buffer views make the two policies compose.
On the lower-bandwidth inter-supernode tier, aggregation reduces launch
and message-count overhead in the evaluated configuration.

\begin{figure*}[t]
\centering
\begin{tikzpicture}[x=1cm,y=1cm]
  % ================= (a) baseline =================
  \node[font=\footnotesize\bfseries, text=hpred, anchor=west]
    at (-8.7,1.15) {(a) Evaluated PyTorch FSDP2 configuration: fused units
    + same-layer RS/AR chain};
  \foreach \lane/\y in {Compute/0.45, AG/-0.4, RS/-1.25, AR/-2.25}
    \node[flabel, anchor=west] at (-8.7,\y) {\lane};
  \foreach \x/\l in {-5.9/N, -2.1/{N-1}, 1.7/{N-2}} {
    \node[ghost, text width=3.15cm, minimum height=22pt] at (\x,0.45)
      {\scriptsize Bwd $L_{\l}$ + grad copyin};
    \node[ghost, text width=3.15cm, minimum height=22pt] at (\x,-0.4)
      {\scriptsize AG $L_{\l}$ + copyout};
    \node[fbox=hpgray, text width=3.15cm, minimum height=22pt]
      (rsa\l) at (\x,-1.25) {\scriptsize RS $L_{\l}$ (fused)};
    \node[bad, text width=3.15cm, minimum height=22pt]
      (ara\l) at (\x,-2.25) {\scriptsize AR $L_{\l}$};
    \draw[-{Stealth[length=1.8mm]}, thick, draw=hpred]
      (rsa\l.south) -- (ara\l.north);
  }
  \node at (5.0,0.45) {$\cdots$}; \node at (5.0,-0.4) {$\cdots$};
  \node at (5.0,-1.25) {$\cdots$}; \node at (5.0,-2.1) {$\cdots$};
  \node[fnote, text=hpred, anchor=west, text width=2.7cm] at (6.0,-0.8)
    {\scriptsize Copy kernels ride the compute stream; AR chained to
     same-layer RS; per-layer AR events re-checked at many points: slow
     tier leaks into backward};
  % ================= (b) ours =================
  \node[font=\footnotesize\bfseries, text=hpgreen, anchor=west]
    at (-8.7,-2.9) {(b) \sys: four-step post-backward pipeline; the slow
    tier is issue-only and settles once at the end of backward};
  \foreach \lane/\y in {Compute/-3.6, AG/-4.45, RS/-5.3, AR/-6.15}
    \node[flabel, anchor=west] at (-8.7,\y) {\lane};
  \foreach \x/\l in {-6.4/N, -3.1/{N-1}, 0.2/{N-2}} {
    \node[good, text width=1.6cm, minimum height=22pt] at (\x-0.2,-3.6)
      {\scriptsize Bwd $L_{\l}$};
    \node[fbox, text width=2.0cm, minimum height=22pt] at (\x,-4.45)
      {\scriptsize AG $L_{\l}$ per-param};
    \node[fbox, text width=2.0cm, minimum height=22pt] (rsb\l) at (\x,-5.3)
      {\scriptsize RS $L_{\l}$ zero-copy};
  }
  \foreach \x/\l in {-5.15/N, -1.85/{N-1}, 1.45/{N-2}}
    \node[fbox=hporange, minimum width=0.85cm, minimum height=22pt]
      (pb\l) at (\x,-3.6) {\scriptsize PB$_{\l}$};
  \node[fbox=hporange, dashed, text width=2.0cm, minimum height=22pt]
    (arbN) at (-3.1,-6.15) {\scriptsize Fused AR $L_N$};
  \node[fbox=hporange, dashed, text width=2.0cm, minimum height=22pt]
    (arbNm) at (0.2,-6.15) {\scriptsize Fused AR $L_{N-1}$};
  \node at (2.6,-3.6) {$\cdots$}; \node at (2.6,-4.45) {$\cdots$};
  \node at (2.6,-5.3) {$\cdots$}; \node at (2.6,-6.15) {$\cdots$};
  \node[good, text width=3.3cm, minimum height=22pt] (root) at (5.9,-3.6)
    {\scriptsize End-of-backward settlement: wait-all ARs, one write-back
     pass};
  \draw[gflow] (pbN.south) -- (rsbN.north);
  \draw[oflow, dashed] (pbN-1.south) -- (arbN.north);
  \draw[oflow, dashed] (pbN-2.south) -- (arbNm.north);
  \draw[oflow, dashed] (arbNm.east) -| (root.south);
  \node[fnote, anchor=west, text width=3.4cm] at (4.1,-5.35)
    {\scriptsize RS of $L_\ell$ overlaps bwd of $L_{\ell-1}$; AR issued one
     layer late, never waited on mid-backward};
  % time axis
  \draw[flow] (-8.7,-7.05) -- (8.4,-7.05);
  \node[fnote] at (0,-7.35) {Time (backward of $L_N \to L_{N-1} \to \cdots$
    $\to$ end of backward)};
\end{tikzpicture}
\caption{Backward communication timelines. (a)~The evaluated PyTorch 2.9
FSDP2 configuration chains
each layer's all-reduce to its own reduce-scatter and surrounds collectives
with copy kernels, so slow inter-supernode latency enters the backward
path. (b)~\sys issues per-parameter reduce-scatter at each layer's
post-backward step and defers the fused all-reduce asynchronously by one
layer; nothing waits on the slow tier until a single end-of-backward
settlement. (PB: post-backward step; AR: all-reduce.)}
\Description{Two backward timelines: the evaluated FSDP2 configuration chaining same-layer reduce-scatter and all-reduce with copy kernels, versus the four-step post-backward pipeline where the slow all-reduce is issue-only and settles once at the end of the backward pass.}
\label{fig:pipeline}
\end{figure*}

\subsection{Four-Step Backward Communication Pipeline}
\label{sec:pipeline}

The backward pass must overlap these collectives without layer-local waits
on the slow tier. \sys schedules it with a deadline-driven argument: a layer's
reduce-scatter has a deadline one layer of backward computation away (the
optimizer will need it soon), while the fused all-reduce's deadline is the
end of the \emph{whole} backward pass. Fast communication therefore goes
first; slow communication consumes the slack.

Concretely, each layer's post-backward step performs exactly four actions
(Algorithm~\ref{alg:pipeline}, Figure~\ref{fig:pipeline}): (1)~\emph{wait} for the previously completed layer's
reduce-scatter, which has already overlapped with this layer's backward
computation and is approximately free; (2)~\emph{settle} gradients of
parameters that need no cross-replica reduction; (3)~\emph{issue} this
layer's reduce-scatter with its output written directly into the
corresponding fused-buffer view; and
(4)~\emph{asynchronously issue} the previously completed layer's fused all-reduce,
whose completion handle is parked in a global pending queue and never
waited on here. At the end of the backward pass, a single settlement
performs the wait-all and writes all final gradients back in one pass. The backward
path therefore contains no layer-local wait on the slow inter-supernode
tier; the end-of-backward settlement enters the step's critical path only
if the
slowest all-reduce outlasts the entire backward hiding window. By contrast,
the evaluated PyTorch 2.9 FSDP2 configuration of \S\ref{sec:setup} issues
the all-reduce immediately after the same layer's reduce-scatter and
maintains per-layer completion events that are
re-checked at many points, letting slow-traffic latency leak into the
backward path.

\begin{algorithm}[t]
\caption{Post-backward step of layer $\ell$ (four-step pipeline; layer
indices in forward order, backward runs $N \to 1$)}
\label{alg:pipeline}
\begin{algorithmic}[1]
\State \textbf{wait} $\mathrm{RS}_{\ell+1}$ \Comment{overlapped during layer $\ell$'s backward}
\State \textbf{settle} gradients of no-reduction parameters of layer $\ell$
\State \textbf{issue} $\mathrm{RS}_{\ell}$ into fused-buffer views \Comment{zero-copy}
\State \textbf{issue-async} $\mathrm{AR}_{\ell+1}$; park handle in pending queue
\Statex \Comment{at end of backward: wait-all pending ARs; single write-back pass}
\end{algorithmic}
\end{algorithm}

\subsection{Interaction with TP/CP/EP Communication Domains}
\label{sec:coexist}

In hybrid parallelism, the backward path mixes communication domains of very
different speeds: tensor-parallel all-gather and reduce-scatter and
expert/context-parallel all-to-all run within the supernode, while the
fused all-reduce crosses supernodes and is the slowest communication in the
system. Any backward-path synchronization on the slow reduction forces the
fast collectives issued synchronously on the compute stream to queue
behind it, so slow traffic leaks into the critical path and is amplified
layer by layer. The schedule of \S\ref{sec:pipeline} removes every
layer-local wait on the slow tier from the backward path by construction
(issue-only, with a single end-of-backward settlement), so the fast domains
observe it only through that final settlement.

\subsection{Engineering Capabilities}
\label{sec:fsdp-eng}

The mechanism above composes with the practical requirements of production
training: uneven (ceil-chunk) sharding on the leading dimension; strided
composition with pre-existing tensor-parallel layouts; mixed precision with
fp32 master gradients; bucket-granularity gradient accumulation; CPU
offloading; recompute-aware prefetch suppression and explicit prefetch
control; and a dual PyTorch/MindSpore backend behind the platform
abstraction of \S\ref{sec:overview}.
      % topology-aware FSDP2/HSDP

\section{Layout-Driven Distributed Muon}
\label{sect:muon}

Muon replaces the element-wise second-moment scaling of AdamW with an
orthogonalization of the momentum matrix by NS iteration,
and has been adopted by a growing list of frontier models
(\S\ref{sec:bg-muon}). This section describes how \sys scales Muon to
multi-dimensional sharded layouts. We first state why the interaction of
matrix semantics with sharding is hard (\S\ref{sec:muon-challenge}) and the
design stance that shapes our answer (\S\ref{sec:philosophy}):
\S\ref{sec:grouping}\textasciitilde\S\ref{sec:muon-pipeline} then present its
components: Grouping parameters from
layout semantics (\S\ref{sec:grouping}), deduplicating orthogonalization
across shard domains (\S\ref{sec:shard-dedup}) and replica domains
(\S\ref{sec:replica-dedup}), batching the NS iteration by core shape
(\S\ref{sec:batched-ns}), and pipelining the whole optimizer step
(\S\ref{sec:muon-pipeline}).

\subsection{Challenges: Matrix Semantics vs.\ Sharded Layouts}
\label{sec:muon-challenge}

Orthogonalization is intrinsically whole-matrix: the Gram factor
$G^{\top}G$ couples the columns of the momentum matrix $G$ in their
entirety (and $GG^{\top}$ the rows), so no shard of
$G$ can be orthogonalized in isolation. Sharding, however, cuts the matrix
apart, and the naive gather-then-compute on every rank remedy repeats
the NS computation $S$-fold across an $S$-way shard group, multiplied a
further $R$-fold across $R$ replicas under hybrid sharding. Worse, the
gather, the NS iteration, and the result return are all new critical-path
work inside the optimizer step, and per-matrix execution is host-bound for
the many small matrices a transformer contains. An all-to-all-based
realization is theoretically redundancy-free, but in our target deployment
it creates an unfavorable connection and resource footprint on the network
interfaces. This deployment constraint motivates our choice rather than
a general conclusion; we therefore choose a gather-based organization in
this deployment regime.

Existing distributed realizations restore whole-matrix optimizer inputs
through materially different mechanisms: Moonlight's bucketed
ZeRO-1-style assignment, DMuon's explicit per-parameter owner routing,
MatrixFSDP's owner-shaped ZeRO-3 placement, Canzona's strategy-specific
static partitioning, and FlexShard's storage-to-compute resharding. They
differ in when communication occurs, where optimizer state resides, and
how computation is scheduled. Against these realizations
(Table~\ref{tab:muonfeat}, Table~\ref{tab:flexshard}; versions as
inspected, runtime revisions pinned in \S\ref{sec:eval-muon}), \sys 
combines three properties in one
design: (i)~communication domains and redundancy derived automatically
from distribution-layout semantics on general multi-dimensional meshes,
including a case that requires no Muon-specific optimizer communication
over the expert-sharding axis when no sharding axis intersects the
matrix plane; (ii)~replica-domain dedup within topology-aligned
subgroups, whose size is determined from the local device count and is 
machine-aligned under the standard contiguous rank
assignment (\S\ref{sec:replica-dedup}), with deterministic
recomputation across subgroups, so that under the
canonical mesh-to-topology mapping (Figure~\ref{fig:topo}) the evaluated
configuration sends no optimizer traffic over the slow interconnect tier;
and (iii)~shape-fused batched orthogonalization inside a local-tensor
programming model backed by one static plan. DMuon already provides
batched orthogonalization and placement-aware tensor-parallel handling,
and computes each matrix exactly once globally; its owner domains are
assigned over a fixed data-parallel mesh rather than derived from layout
semantics, and its routing spans both interconnect tiers. MatrixFSDP
places each whole matrix on one data-parallel rank (its \emph{parameter
owner}, holding the matrix and its optimizer state) under ZeRO-3
sharding, so its optimizer step issues no matrix collective; the owner
placement is specific to the data-parallel plane and excludes
tensor-parallel-fragmented matrices. Canzona statically assigns whole
parameters to data-parallel ranks for communication-free updates and
batches tensor-parallel fragments into asynchronous intra-node
micro-groups; its domains are fixed per parallel strategy rather than
derived from general layout semantics. TorchTitan's
FlexShard, a general storage-to-compute redistribution substrate, is
contrasted in \S\ref{sec:philosophy}.

\begin{table*}[t]
\caption{Design contrast of distributed Muon realizations (versions as
inspected, runtime revisions pinned in \S\ref{sec:eval-muon}): the six
dimensions most relevant to \sys's
design stance, against DMuon~\cite{dmuon}, MatrixFSDP~\cite{matrixfsdp},
Canzona~\cite{canzona}, and TorchTitan's
FlexShard/DistMuon~\cite{torchtitan,flexshard}. The full eleven-dimensional
comparison matrix is in Appendix~\ref{app:muon-matrix}.}
\label{tab:flexshard}
{\scriptsize
\setlength{\tabcolsep}{3pt}
\begin{tabular}{@{}>{\RaggedRight\arraybackslash}p{1.8cm}>{\RaggedRight\arraybackslash}p{2.95cm}>{\RaggedRight\arraybackslash}p{2.95cm}>{\RaggedRight\arraybackslash}p{2.95cm}>{\RaggedRight\arraybackslash}p{2.95cm}>{\RaggedRight\arraybackslash}p{2.95cm}@{}}
\toprule
 & \textbf{\sys Muon} & \textbf{DMuon} & \textbf{MatrixFSDP} & \textbf{Canzona} & \textbf{FlexShard DistMuon} \\
\midrule
\textbf{Domain derivation} & From layout semantics on general meshes (matrix-plane predicate) & Owner assignment over the DP mesh; nested TP ownership from placements & Owner-shaped ZeRO-3 shards: one whole-matrix owner per 2-D weight & Logical assignment decoupled from physical layout; static per-strategy partition & User-declared compute layouts on named DeviceMesh axes \\ \midrule
\textbf{State dedup scope} & Shard + replica two-level; replica state reduced to $1/R_s$ in topology-aligned subgroups & One authoritative owner per matrix (parameter + momentum) & Whole matrix and state on the owner; empty shards elsewhere & Whole-parameter state at the statically assigned rank & Owner-style state on the compute layout (\emph{Owned}, \emph{BlockShard}) \\ \midrule
\textbf{Cross-topology strategy} & Replica subgroups from the local device count (contiguous rank blocks, machine-aligned under the standard rank mapping; a configuration premise, \S\ref{sec:replica-dedup}); deterministic recomputation; no slow-tier optimizer traffic under the canonical mapping (Fig.~\ref{fig:topo}) & Two-stage intra-/inter-node hierarchy with XOR owner slots; spans both tiers & Optimizer step local; the backward reduction spans the DP mesh & DP step zero-communication; TP reconstruction kept intra-node & Topology from the user-chosen mesh and layouts; no automatic machine-boundary policy described \\ \midrule
\textbf{Optimizer-step collectives} & Fused all-gather + relay broadcast; none over the expert-sharding axis when the matrix plane is unsharded & Owner-to-all broadcast / all-to-owner reduce & None. The backward reduce-scatter lands the input on the owner & DP: none; TP: fused all-to-all micro-groups & Packed all-to-all redistribution \\ \midrule
\textbf{Batching / Load balance} & Shape-fused batched NS; greedy owner balance; ${\le}1$ batch in flight & Gram-NS batching with SYRK kernels and autotuning; measured (MILP) owner assignment & Global owner planner balancing resident bytes and optimizer work & $\alpha$-balanced LPT (DP); micro-group scheduling with rollback (TP) & \emph{BucketConfig} grouping and ordering for packed-redistribution overlap \\ \midrule
\textbf{Mesh coverage} & General multi-dimensional meshes; expert-axis case needs no Muon-specific optimizer communication & FSDP2/HSDP data-parallel mesh + nested TP & ZeRO-3 data parallelism; TP-fragmented matrices excluded & ZeRO-1 DP + TP; optimizer-agnostic (Muon, Shampoo, SOAP) & Layouts on named mesh axes; DistMuon flat matrix-batch: \emph{BlockShard} on one non-unit axis only \\
\bottomrule
\end{tabular}}
\end{table*}

\subsection{Design Stance: a Local Programming Model over an Internal Global Plan}
\label{sec:philosophy}

The organizing principle of \sys's Muon is that distribution is an
execution concern, not an algorithmic one. The algorithm is written
against \emph{local tensors}: each rank sees only its gradients, its
momentum, the matrices it must orthogonalize, the Newton-Schulz
iteration, and the parameter update. The distribution layer answers two
questions only: How each parameter is currently sharded, and which
process groups that sharding spans. It is never asked to re-express the
optimizer's internal decisions (which rank computes a given matrix, and
in what order) as a second, optimizer-facing layout system.

The global organization exists, but inside the execution engine rather
than in the algorithm's vocabulary. The ranks joining each collective,
the compute owner of every matrix, per-rank send and receive volumes,
collective and bucket ordering, and buffer sizing and reuse are all fixed
by a static plan when the optimizer is built, and replayed unchanged at
every step. One step traverses: local gradient $\to$ replica-owner
deduplication $\to$ momentum update $\to$ fused all-gather $\to$
load-balanced compute-owner assignment $\to$ batched NS $\to$ packed
relay broadcast $\to$ slice extraction and local update $\to$ replica
broadcast. An internal global plan; an external local programming model.

Two consequences follow. First, communication primitives follow the
hardware rather than the abstraction. Fused all-gather plus owner
broadcast (i.e., hierarchical relay across mesh axes) matches the fast paths
of collective-communication libraries, fixes collective order statically,
makes buffer sizes and lifetimes pre-plannable, and avoids the rank-pair
queue-pair pressure of all-to-all collectives on the slow tier
(\S\ref{sec:muon-challenge}); the policy can be retargeted per platform
without touching the algorithm interface. Second, the distributed
skeleton is decoupled from the algorithm it carries: the momentum rule,
the orthogonalization itself (step count, coefficients, epsilon), input
reshaping and matrix extraction, RMS-matched scaling, and post-update
processing are all pluggable or configurable stages, so legacy Muon, the asymmetric
five-step variant, per-parameter coefficient sets, alternative momentum
rules, and QK-clip-style post-processing all reuse the same owners,
buckets, buffers, and communication plan.

Table~\ref{tab:flexshard} contrasts this stance with the realizations of
\S\ref{sec:muon-challenge} along the dimensions most relevant to it
(Appendix~\ref{app:muon-matrix} gives the full matrix). The sharpest
contrast is with FlexShard, TorchTitan's general substrate for
distributed Muon~\cite{torchtitan,flexshard}. FlexShard
gives the optimizer a second layout system: a per-parameter
\emph{compute layout} (block-sharded or owner-assigned) is declared next
to the storage layout, and a general reshard planner lowers the
storage-to-compute transition onto packed all-to-all collectives. The
abstraction is complete and serves many optimizers; its costs are a
second layout vocabulary for algorithm developers and an all-to-all
communication profile. DMuon assigns each matrix a single authoritative
owner over the data-parallel mesh and reaches near-AdamW overhead through
hierarchical communication, Gram-space batched Newton-Schulz, and
measured load balancing; its domains are assigned over a fixed mesh
hierarchy rather than derived from layout semantics, and its publication
and reduction traffic spans both interconnect tiers. MatrixFSDP and
Canzona likewise keep a conventional optimizer interface, but their
domains follow fixed owner-shaped ZeRO-3 placement,  respectively a 
static DP/TP assignment, rather than the layout semantics
of a general multi-dimensional mesh. \sys makes a
different trade: specializing to
Muon's fixed structure enables statically pre-plannable communication and
hardware-friendly collectives, at the cost of a general optimizer layout
abstraction. The distributed tensor stays at its most valuable
boundary: Describing how parameters are distributed, not everything the
optimizer then does with them.

\subsection{Layout-Semantics-Driven Parameter Grouping}
\label{sec:grouping}

\sys requires no user-supplied process groups. For every parameter, the
optimizer extracts from the distribution metadata which mesh axes shard it
(along which tensor dimension) and which axes replicate it; one code path
serves both \sys's own distributed tensors and PyTorch-native ones, because
both expose the shard/replicate placement metadata this grouping path
requires.

\begin{figure}[t]
\centering
\begin{tikzpicture}[x=1cm,y=1cm]
  \node[fbox, text width=7.2cm] (p) at (0,0)
    {\textbf{Parameter + Placements} {\scriptsize(either backend)}};
  \node[fbox, text width=7.2cm] (ex) at (0,-0.9)
    {\scriptsize Extract layout (shape-free): which mesh axes shard which
     tensor dims, along which tensor dim; which axes replicate};
  \node[warn, text width=6.2cm, minimum height=30pt] (dec) at (0,-2.0)
    {\scriptsize Does any sharding axis intersect the \textbf{matrix plane}
     (last two dims)?};
  \node[fbox, text width=3.42cm, minimum height=78pt] (ya) at (-1.94,-4.35)
    {\textbf{Yes}\\[1pt]
     {\scriptsize e.g., FSDP-sharded Linear $[out,in]$ on dim 0:
      each rank holds rows only}\\[2pt]
     {\scriptsize $\to$ shard-domain dedup (\S\ref{sec:shard-dedup}):
      fused gather, owner NS, relay, slice}};
  \node[good, text width=3.42cm, minimum height=78pt] (na) at (1.94,-4.35)
    {\textbf{No}\\[1pt]
     {\scriptsize e.g., MoE expert $[E,in,out]$ sharded on $E$: each rank
      holds whole expert matrices}\\[2pt]
     {\scriptsize $\to$ local batched NS (\S\ref{sec:batched-ns}): no
      Muon-specific optimizer communication over the expert-sharding axis,
      no redundancy}};
  \node[ghost, text width=7.2cm] (c) at (0,-7.05)
    {\scriptsize 1-D / scalar parameters $\to$ AdamW (chained optimizer;
     RMS matching transfers the learning-rate scale; decoupled weight
     decay is applied separately)};
  \draw[flow] (p) -- (ex);
  \draw[flow] (ex) -- (dec);
  \draw[flow] (dec.south -| ya.north) -- node[left, flabel] {yes} (ya.north);
  \draw[flow] (dec.south -| na.north) -- node[right, flabel] {no} (na.north);
  \draw[flow] (dec.south) -- (c.north);
\end{tikzpicture}
\caption{Layout-semantics-driven grouping. Communication cost is
determined by which matrices are cut and by the sizes of the cut matrix
planes, not by raw parameter count alone: only
matrix-plane-sharded parameters gather; expert-axis-sharded weights, which are the
most numerous under expert parallelism, require no Muon-specific
optimizer communication over the expert-sharding axis.}
\Description{Decision flow: from parameter placements, extract sharding and replication axes, test whether any sharding axis intersects the matrix plane; yes goes to shard-domain dedup, no to local batched Newton-Schulz without optimizer communication, and 1-D parameters to AdamW.}
\label{fig:muon-grouping}
\end{figure}
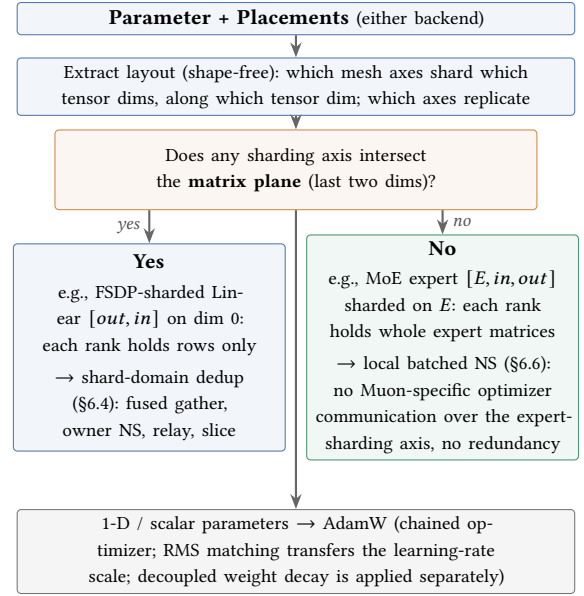

The grouping rests on one predicate (Figure~\ref{fig:muon-grouping}): does
any sharding axis intersect the
\emph{matrix plane}, i.e., the last two tensor dimensions on which NS operates?
If not (the canonical case is the expert axis of a 3-D expert weight, which
slices \emph{between} matrices rather than through them), every locally
held matrix slice is already complete, and orthogonalization is purely
local, i.e., no Muon-specific optimizer communication over the expert-sharding
axis, and is redundancy-free. The governing
principle: communication cost is determined by which matrices are cut and
by the sizes of the cut matrix planes, not by raw parameter count alone. One-dimensional parameters leave Muon
entirely and are delegated to AdamW by a chained optimizer.

The expert case deserves one precision. A fused expert parameter of shape
$[E,M,N]$ denotes $E$ independent $[M,N]$ matrices: Muon must run the NS
iteration per expert matrix, and the three-dimensional parameter must
never be flattened into a single $[E, MN]$ matrix for orthogonalization.

\subsection{Shard-Domain Deduplication}
\label{sec:shard-dedup}

\begin{figure*}[t]
\centering
\begin{tikzpicture}[x=1cm,y=1cm]
  % ---------- (a) shard-domain dedup ----------
  \node[font=\footnotesize\bfseries, text=hpblue, anchor=west]
    at (-8.4,1.05) {(a) Shard Domain: Each matrix orthogonalized exactly
    once per shard group};
  \node[bad, text width=3.0cm, minimum height=64pt] (naive) at (-6.9,-0.5)
    {\textbf{Naive}\\[1pt]
     {\scriptsize Gather, then every rank runs NS over all $S$ matrices:
      $S\times$ redundant compute}};
  \node[fbox, text width=2.55cm, minimum height=64pt] (st1) at (-2.9,-0.5)
    {\textbf{1. Fused Gather}\\[1pt]
     {\scriptsize One collective per sharding axis packs a whole batch
      (bf16, aligned, uneven splits)}};
  \node[fbox, text width=2.55cm, minimum height=64pt] (st2) at (0.15,-0.5)
    {\textbf{2. Owner Assignment}\\[1pt]
     {\scriptsize Greedy size balance; only the owner materializes the full
      matrix and runs NS}};
  \node[fbox, text width=2.55cm, minimum height=64pt] (st3) at (3.2,-0.5)
    {\textbf{3. Relay Broadcast}\\[1pt]
     {\scriptsize Owners pack updates; broadcast per sharding axis,
      asynchronous}};
  \node[fbox, text width=2.55cm, minimum height=64pt] (st4) at (6.25,-0.5)
    {\textbf{4. Slice-Apply}\\[1pt]
     {\scriptsize Each rank extracts its own slice; momentum stays sharded
      throughout}};
  \draw[-{Stealth[length=1.8mm]}, thick, draw=hpred, dashed]
    (naive.east) -- (st1.west);
  \draw[flow] (st1) -- (st2);
  \draw[flow] (st2) -- (st3);
  \draw[flow] (st3) -- (st4);
  % ---------- (b) replica-domain dedup ----------
  \node[font=\footnotesize\bfseries, text=hpblue, anchor=west]
    at (-8.4,-2.35) {(b) Replica Domain: Within each subgroup, one of
    $R_s$ identical copies computes; subgroups sized from the local
    device count};
  \node[ghost, text width=2.15cm, minimum height=52pt] (rp0) at (-7.3,-3.85)
    {\textbf{Replica 0}\\[1pt] {\scriptsize No momentum buffer, no NS}};
  \node[ghost, text width=2.15cm, minimum height=52pt] (rp1) at (-4.75,-3.85)
    {\textbf{Replica 1}\\[1pt] {\scriptsize No momentum buffer, no NS}};
  \node[good, text width=2.15cm, minimum height=52pt] (rp2) at (-2.2,-3.85)
    {\textbf{Replica 2 = owner}\\[1pt] {\scriptsize Momentum + NS + apply:
      computes once}};
  \node[ghost, text width=2.15cm, minimum height=52pt] (rp3) at (0.35,-3.85)
    {\textbf{Replica 3}\\[1pt] {\scriptsize No momentum buffer, no NS}};
  \draw[-{Stealth[length=2.4mm]}, line width=1.1pt, draw=hpgreen!70!black]
    (rp2.north) to[bend left=18] (rp0.north);
  \draw[-{Stealth[length=2.4mm]}, line width=1.1pt, draw=hpgreen!70!black]
    (rp2.north) -- (rp1.north);
  \draw[-{Stealth[length=2.4mm]}, line width=1.1pt, draw=hpgreen!70!black]
    (rp2.north) to[bend right=18] (rp3.north);
  \node[fnote, fill=white, inner sep=1.5pt] at (-2.2,-2.62)
    {\scriptsize Updated-parameter broadcast};
  \node[fnote, anchor=west, text width=5.6cm] at (2.1,-3.85)
    {\scriptsize Updated parameters broadcast hierarchically, one replica
     axis at a time; Muon state of eligible matrix parameters reduced to
     $1/R_s$ ($R_s$: subgroup size);
     every matrix computed once per topology-aligned replica subgroup};
\end{tikzpicture}
\caption{Two-level deduplication of orthogonalization.
(a)~Shard domain: a fused gather feeds greedy owner assignment, so each
matrix is orthogonalized exactly once per shard group instead of once per
rank. (b)~Replica domain: only owner ranks hold momentum and compute; the
updated parameters are broadcast hierarchically, and replica groups are
split into contiguous rank blocks sized from the local device count
(aligned with machine boundaries under the standard rank assignment), so
replica-domain traffic stays on the fast tier; shard-domain collectives
stay intra-supernode under the canonical mapping
(Figure~\ref{fig:topo}).}
\Description{Two-part diagram of deduplication: (a) shard domain with fused gather, greedy owner assignment, relay broadcast, and slice-apply so each matrix is orthogonalized once per shard group; (b) replica domain where only the owner computes and replica subgroups are contiguous rank blocks sized from the local device count, aligned with machine boundaries under the standard rank mapping as a configuration premise.}
\label{fig:muon-dedup}
\end{figure*}
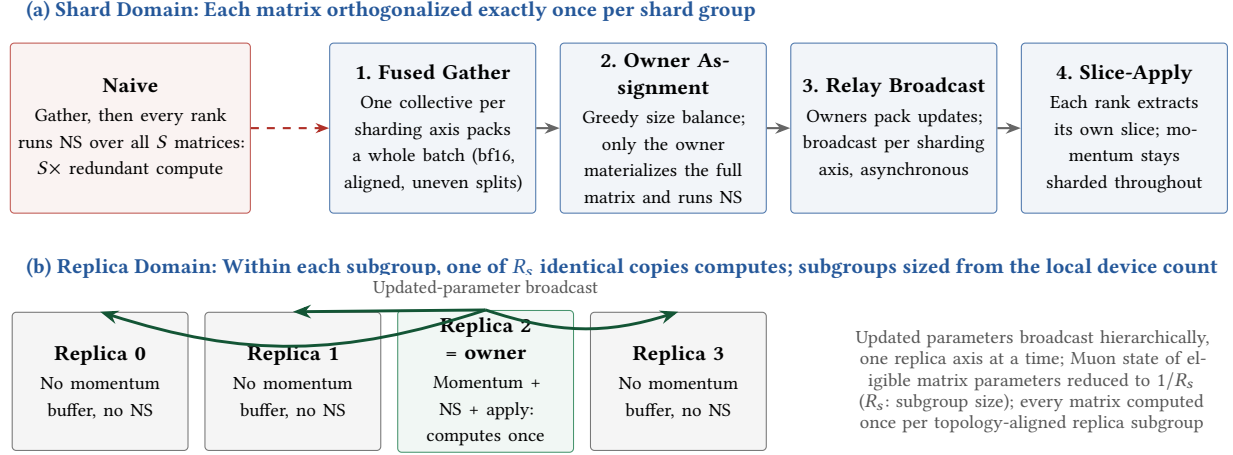

For parameters whose matrix plane \emph{is} sharded, \sys eliminates the
$S$-fold redundancy in four steps (Figure~\ref{fig:muon-dedup}(a)). (1)~\emph{Fused gather}: one collective
per sharding axis packs a whole batch of momentum shards into an aligned
staging buffer (bf16, uneven splits supported). (2)~\emph{Owner
assignment}: each matrix is assigned to exactly one rank of the shard group
by greedy size balancing (its \emph{compute owner}); only the owner
materializes the full matrix and
runs NS, so each matrix is orthogonalized \emph{exactly once} per shard
group while all matrices of a batch still progress in parallel across the
group. (3)~\emph{Relay broadcast}: owners pack their updates and broadcast
them asynchronously, dimension by dimension, along the sharding axes.
(4)~\emph{Slice and apply}: each rank extracts its own slice and applies
the update with weight decay through fused elementwise kernels. Momentum
stays sharded throughout, so optimizer-state memory never reflows.

\subsection{Replica-Domain Deduplication}
\label{sec:replica-dedup}

Replica axes contribute the second, multiplicative redundancy: $R$
identical copies of every replicated matrix would otherwise be
orthogonalized $R$ times. \sys deduplicates within each
\emph{topology-aligned replica subgroup}
(Figure~\ref{fig:muon-dedup}(b)): momentum buffers exist
only on one \emph{replica-owner} rank per subgroup, only replica owners
orthogonalize, and the updated parameters
are broadcast to peers hierarchically and asynchronously, one replica
dimension at a time. Writing $R_s$ for the subgroup size, the Muon
momentum state of eligible matrix parameters is reduced to $1/R_s$ of
the one-copy-per-replica baseline, and each matrix is
orthogonalized once per subgroup, which is $R/R_s$ times in total across the
original replica group, degenerating to a single global computation when
the group is not split ($R_s = R$).

The dedup domain's topology alignment is a \emph{configuration premise},
not physical-topology probing: the realization derives the subgroup size
from the per-machine device count by selecting the largest divisor of the
replica-group size not exceeding it, with no split at all when the group
already fits on one machine, and partitions each replica group's sorted
ranks into contiguous blocks of that size. This coincides with machine
boundaries under the standard contiguous rank-to-node assignment of the
evaluated launchers. Note that a machine (node) is defined as one physical host inside a
supernode, so the boundary lies strictly inside the supernode's
high-bandwidth domain. A rank permutation that interleaved machines, or
a replica group smaller than one machine yet spanning two, would violate
the premise undetected; making the partition read the physical
rank-to-node mapping (and fail fast when a subgroup would cross a slow
boundary) is implementation future work, and the evaluated
configuration's alignment is checked with the link-tier traffic
measurements of our experiment plan. Within each subgroup, deduplication
and broadcast run \emph{locally},
while across subgroups \sys uses deterministic
\emph{re-computation}. Since gradients have already been reduced identically
(\S\ref{sec:twotier}) and the NS iteration is deterministic, every
subgroup derives numerically equivalent results with no communication
(identical inputs and a deterministic iteration; bitwise identity
additionally requires a fixed kernel set and software stack, which our
deployments pin). Under this premise, replica-domain traffic never
crosses the slow
tier. Shard-domain collectives inherit the topology of their sharding
axes: under the canonical mesh-to-topology mapping of
Figure~\ref{fig:topo} where sharding axes inside the supernode, replica axes
across supernodes, the collectives remain intra-supernode, so the evaluated
configuration sends no Muon-specific optimizer traffic over the slow
tier. A mapping that placed a sharding axis across machines would need an
analogous bound on the shard domain, which the current realization does
not enforce automatically.

\subsection{Shape-Fused Batched Orthogonalization}
\label{sec:batched-ns}

\begin{figure*}[t]
\centering
\begin{tikzpicture}[x=1cm,y=1cm]
  % ---------- (a) shape-fused batched NS ----------
  \node[font=\footnotesize\bfseries, text=hpblue, anchor=west]
    at (-8.4,0.95) {(a) Shape Fusion: One batched kernel per core-shape
    group};
  \node[ghost, text width=2.0cm, minimum height=64pt] (ps) at (-7.1,-0.6)
    {\scriptsize $p_1$ $[A,B]$\\[2pt] $p_2$ $[A,B]$\\[2pt]
     $p_3$ $[N,A,B]$\\[2pt] $p_4$ $[A,1,B]$};
  \node[fbox, text width=1.9cm, minimum height=64pt] (cat) at (-4.35,-0.6)
    {\scriptsize Concat on batch axis\\[2pt] Same core shape $(A,B)$};
  \node[fbox, text width=1.55cm, minimum height=64pt] (merged) at (-1.85,-0.6)
    {\scriptsize Merged $[K,A,B]$};
  \node[good, text width=2.3cm, minimum height=64pt] (bns) at (0.6,-0.6)
    {\scriptsize \textbf{One Batched NS}: 5 steps, batched matmuls};
  \node[fbox, text width=1.85cm, minimum height=64pt] (split) at (3.15,-0.6)
    {\scriptsize Split back; apply per-parameter updates};
  \draw[flow] (ps) -- (cat);
  \draw[flow] (cat) -- (merged);
  \draw[flow] (merged) -- (bns);
  \draw[flow] (bns) -- (split);
  \node[fnote, anchor=west, text width=3.6cm] at (4.9,-0.6)
    {\scriptsize bf16, workspace allocated once per batched invocation;
     transpose-first when
     rows$>$cols; reshape hook for fused QKV; cap $2^{29}$ elements scaled
     by shard-group size};
  % ---------- (b) batch-level pipelining ----------
  \node[font=\footnotesize\bfseries, text=hpblue, anchor=west]
    at (-8.4,-2.35) {(b) Batch-Level Pipelining: Broadcasts sink beneath
    compute};
  \node[flabel, anchor=west] at (-8.5,-3.15) {Compute};
  \node[flabel, anchor=west] at (-8.5,-3.95) {Comm.};
  \node[fbox, text width=2.2cm, minimum height=26pt, align=center]
    (b1) at (-6.3,-3.15) {\scriptsize Batch 1: NS (largest)};
  \node[fbox, text width=2.2cm, minimum height=26pt, align=center]
    (b2) at (-3.5,-3.15) {\scriptsize Batch 2: NS};
  \node[fbox, text width=2.2cm, minimum height=26pt, align=center]
    (b3) at (-0.7,-3.15) {\scriptsize Batch 3: NS};
  \node[fbox=hporange, dashed, text width=2.2cm, minimum height=26pt,
    align=center] (bc1) at (-3.5,-4.25) {\scriptsize Broadcast batch 1
    (async)};
  \node[fbox=hporange, dashed, text width=2.2cm, minimum height=26pt,
    align=center] (bc2) at (-0.7,-4.25) {\scriptsize Broadcast batch 2};
  \node[good, text width=1.75cm, minimum height=26pt] (wait) at (2.15,-3.15)
    {\scriptsize Final drain: wait all};
  \node at (3.55,-3.15) {$\cdots$};
  \draw[oflow, dashed] (b1.south) -- (bc1.north);
  \draw[oflow, dashed] (b2.south) -- (bc2.north);
  \node[fnote, anchor=west, text width=4.3cm] at (4.3,-3.55)
    {\scriptsize Batches packed in descending size; at most one batch in
     flight; admitting a new batch settles the oldest; per-axis relay
     removes host-side bubbles};
  \draw[flow] (-8.4,-4.8) -- (8.4,-4.8);
  \node[fnote] at (0,-5.08) {Time};
\end{tikzpicture}
\caption{Batched orthogonalization and pipelining. (a)~Parameters sharing a
core matrix shape ($[A,B]$, $[A,1,B]$, $[N,A,B]$) are concatenated and
orthogonalized by one batched Newton-Schulz kernel. (b)~Each batch's
broadcast is issued asynchronously as soon as the batch finishes, so the
timeline degenerates to mostly continuous compute; the bounded in-flight
queue may settle an earlier batch before admitting the next one, and the
final drain waits for all remaining work.}
\Description{Two-part diagram: (a) parameters sharing a core matrix shape are concatenated and orthogonalized by one batched Newton-Schulz kernel; (b) batch-level pipelining with asynchronous broadcasts, at most one batch in flight, and a final drain.}
\label{fig:muon-batch}
\end{figure*}
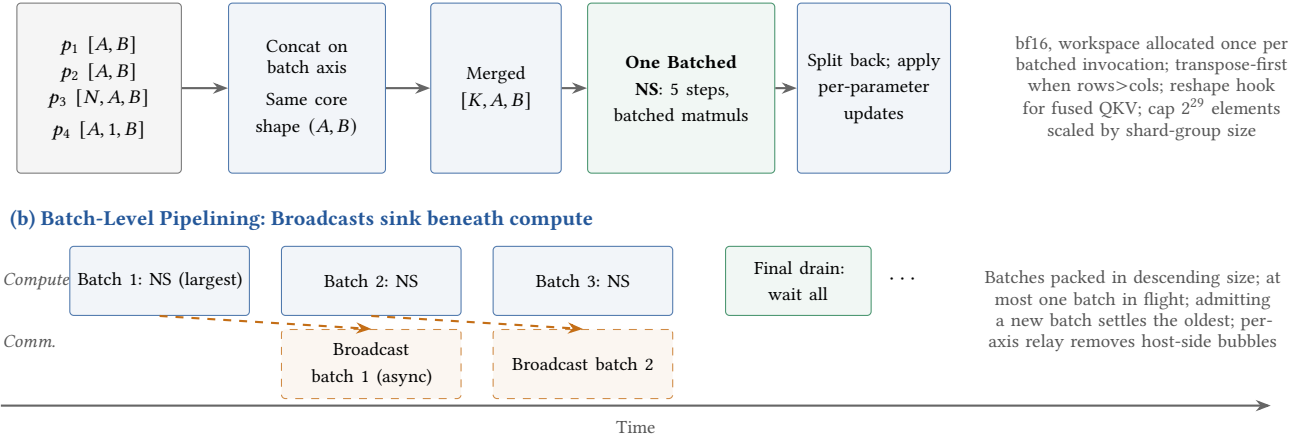

Per-matrix NS launches starve the device for small matrices. \sys groups
parameters by their \emph{core matrix shape}
(Figure~\ref{fig:muon-batch}(a)): tensors of shape $[A,B]$,
$[A,1,B]$, and $[N,A,B]$ share the $(A,B)$ group, are concatenated along a
batch axis, orthogonalized by \emph{one} batched NS iteration (a single
sequence of batched matmuls), and split back.

Parameter storage layout and Muon's mathematical matrices are
deliberately decoupled. One physical parameter can hold several logical
matrices, e.g., the heads of a fused QKV projection, the gate and up matrices
packed into an MoE FC1 weight, the two query factors of an MLA
projection, or the key and value factors of its KV-up projection. Rather
than pushing these model semantics into the distribution layout, \sys
defines \emph{the matrix as Muon sees it} in the model adapter: a
zero-copy view reshape, an arbitrary reversible
split/concatenate/interleave transform applied to the NS input, and a
restore mapping that writes the logical-matrix updates back into the
physical parameter. Model-structure changes thereby stay in the model
adapter and never propagate into placement metadata, the sharding
planner, or the communication runtime.

The kernels are engineered for the regime: bf16 arithmetic with
workspace allocated once per batched invocation rather than per matrix; a transpose-first rule when rows exceed columns so the Gram
product falls on the small dimension; two selectable coefficient sets (the
legacy quintic and an asymmetric five-step variant); and memory-safe
batching capped at $2^{29}$ elements, scaled down by the shard-group size.
For the many small matrices a production MoE contains, batching replaces
dozens of host-bound per-matrix launches with a single batched kernel.

\subsection{Batch-Level Pipelining}
\label{sec:muon-pipeline}

\begin{algorithm}[t]
\caption{One distributed Muon step}
\label{alg:muon-step}
\begin{algorithmic}[1]
\State partition parameters by the matrix-plane predicate
  (\S\ref{sec:grouping})
\State \textbf{complete path:} momentum update $\to$ batched NS $\to$
  apply
\Statex \hspace{\algorithmicindent} \Comment{no Muon-specific optimizer
  communication over the expert-sharding axis}
\ForAll{batches $B$ of matrix-plane-sharded params., descending size}
  \State fused-gather momentum shards of $B$ per sharding axis
  \State assign matrices of $B$ to owners (greedy size balance)
  \State owners: batched NS over core-shape groups
  \State issue-async relay broadcast of $B$ along shard axes
  \State subgroup shard ranks: slice from relay; apply with decay
  \State replica owners: issue-async per-axis broadcast
  \Statex \hspace{\algorithmicindent} \Comment{$\le 1$ in flight; new
  batch settles oldest}
\EndFor
\State \textbf{final drain:} wait all outstanding broadcasts
\end{algorithmic}
\end{algorithm}

Finally, the optimizer step itself is pipelined
(Figure~\ref{fig:muon-batch}(b)). Parameters are packed into
batches in descending size order (i.e., large matrices first) so their result
broadcasts enjoy the longest hiding window. Each batch's replica
broadcast is issued asynchronously as soon as the batch finishes, with at
most one batch in flight. Admitting a new batch settles the oldest, so an
intermediate batch may wait briefly at that point;
dimension-by-dimension relay removes host-side
bubbles. The step timeline degenerates to a continuous stream of momentum
and orthogonalization compute with communication sunk beneath it; a final
drain waits for all remaining work (Algorithm~\ref{alg:muon-step}).
      % distributed Muon

\begin{table*}[t]
\caption{Evaluated configurations. A: dual-mode tensor-runtime study
(production vs.\ validation mode at MBS 1 and 2 under one sharding plan;
\S\ref{sec:eval-dtensor}). B: 16-die pure-FSDP comparison against PyTorch
FSDP2 (\S\ref{sec:eval-fsdp}). C: 128-rank dense-model comparison against
Megatron DDP with distributed optimizer (\S\ref{sec:eval-fsdp}); the
baseline column of B/C is described in the text. D: 505B-parameter MoE
scale demonstration without a baseline (\S\ref{sec:eval-fsdp}). Parallel
meshes describe the \sys side.}
\label{tab:setup}
{\scriptsize
\setlength{\tabcolsep}{2.5pt}
\begin{tabular}{@{}>{\RaggedRight\arraybackslash}p{1.55cm}>{\RaggedRight\arraybackslash}p{3.8cm}>{\RaggedRight\arraybackslash}p{3.8cm}>{\RaggedRight\arraybackslash}p{3.8cm}>{\RaggedRight\arraybackslash}p{3.8cm}@{}}
\toprule
\textbf{Field} & \textbf{A: dual-mode} & \textbf{B: 16-die FSDP} & \textbf{C: 128-rank dense} & \textbf{D: 505B MoE} \\
\midrule
\textbf{Model} & Qwen3-30B-A3B (48 layers, 128 experts, top-8) & Qwen3-30B-A3B (same as A) & 7BV2 dense (39 layers, GQA, SWA, MTP-4) & 505B MoE (51 layers, MLA, 384 experts, top-8) \\ \midrule
\textbf{Hardware} & 16 Ascend NPU devices (16 ranks) & Ascend A3 system, 16 NPU dies & 8 nodes $\times$ 16 dies $=$ 128 ranks, Ascend 910C & 48 nodes, 384 Ascend 910C cards $\times$ 2 dies $=$ 768 ranks \\ \midrule
\textbf{Parallelism} (\sys) & TP4$\cdot$EP4$\cdot$CP1; DP-shard 4; EDP-shard 4 & pure FSDP, shard 16 (TP$=$EP$=$CP$=$PP$=$1); reshard after fwd/bwd; prefetch depth 1 & TP4$\cdot$DP32; HSDP 16-shard $\times$ 2-replica; reshard-after-forward; prefetch depth 2 & TP4$\cdot$EP16$\cdot$DP192$\cdot$PP1; non-expert shard 32 $\times$ 6 replicas; expert shard 6 $\times$ 2 replicas; prefetch depth 2 \\ \midrule
\textbf{Workload} & seq 4096; GBS 32; MBS 1 and 2; 131{,}072 tok/step & seq 4096; GBS 32; MBS 2; 131{,}072 tok/step & seq 4096; GBS 384; MBS 4; 3 grad-accum; 1{,}572{,}864 tok/step & seq 4096; GBS 768; MBS 4; 3{,}145{,}728 tok/step \\ \midrule
\textbf{Precision} & bf16 model dtype; mixed-precision policy off & bf16 parameters; HCCL bf16 & bf16 compute; fp32 softmax/logits/gradient reduction & bf16 parameters; fp32 master gradients and reduction \\ \midrule
\textbf{Recompute} & off & full, layers 0\textasciitilde 26 & off (both systems) & full/block, 47 of 51 layers \\ \midrule
Optimizer & Muon + AdamW groups & Muon + AdamW groups & Muon (both systems) & Muon (all-gather strategy) + AdamW groups \\ \midrule
\textbf{Software} & torch 2.9.0+cpu; torch\_npu 2.9.0; CANN 8.5.1; Transformers 5.13.0; \sys at \texttt{bada2df4} with the archived working-tree patch & torch\_npu 2.9.0; CANN 8.5.1 & PyTorch 2.6.0; torch\_npu 2.6.0.post5; CANN 8.5.0 & PyTorch 2.6.0; torch\_npu 2.6.0.post5; CANN 8.5.2 \\ \midrule
\textbf{Measurement} & two mode pairs, each 60 steps; window steps 2\textasciitilde 60 excluding profiler-export step 13 ($n=58$ per run) & 60 steps; window steps 8\textasciitilde 60 ($n=53$); single-step rank-0 profiler & 1{,}000 steps; window steps 100\textasciitilde 999 ($n=900$); one run per system & window steps 30\textasciitilde 50 ($n=21$); no baseline \\
\bottomrule
\end{tabular}}
\end{table*}

\section{Evaluation}
\label{sect:eval}

We evaluate \sys on Ascend hardware at three scales: a 16-device system,
a 128-rank cluster, and a 384-card (768-rank) cluster. The evaluation is organized by the
three design threads: \emph{dual-mode distributed-tensor execution}
(\S\ref{sec:eval-dtensor}), \emph{topology-aware fully-sharded parallelism}
(\S\ref{sec:eval-fsdp}), and \emph{layout-driven distributed Muon}
(\S\ref{sec:eval-muon}). The tensor-runtime study runs the production and
validation modes of one sharding plan at two micro-batch sizes and compares
device work, memory, numerics, and step time. The fully-sharded study
reports two whole-system comparisons against PyTorch FSDP2 at 16 dies
and against Megatron DDP with distributed optimizer at 128 ranks, plus one
baseline-free scale demonstration on a 505B-parameter MoE. The Muon study
compares three baseline integrations at 16 NPUs and discloses the
implementation issues the comparisons surfaced. We do not perform
per-component ablations and do not attribute whole-system differences to
individual mechanisms; where the compared configurations differ, the delta
is disclosed next to the result.

\subsection{Experimental Setup}
\label{sec:setup}

Table~\ref{tab:setup} fixes the four evaluated configurations; the
subsections below add per-experiment detail and disclose baseline-side
configuration deltas next to each result. All experiments run on Atlas 900 A3 SuperPoD with HCCL collectives.

\paragraph{MFU Accounting.} All reported MFU figures use one shared
accounting rule with coefficients 1 for forward computation and 2 for
backward computation: causal attention of sequence length $L$ is counted over
the $L(L{+}1)/2$ effective query--key positions of the lower triangle, not
$L^2$; MoE layers are counted over the FLOPs of the \emph{activated}
(routed) experts rather than all experts. MFU is reported for experiments
C and D only: in experiment C both systems derive MFU from the same
per-chip peak denominator ($\approx$353 TFLOP/s, recovered from the logs'
TFLOP/s-to-MFU ratio), so the two values are directly comparable;
experiment D quotes its training log's own MFU accounting. Experiments A
and B report step time and shape-derived nominal throughput instead.

\begin{table*}[t]
\caption{Production and validation performance at two micro-batch sizes.
Parentheses show the validation-mode change relative to production at the
same MBS.  Throughput is nominal, computed from 131{,}072 tokens per step.
The primary window is steps 2\textasciitilde 60 excluding profiler-export step 13
($n=58$ per run).}
\label{tab:dual-mode}
\centering
{\small
\setlength{\tabcolsep}{6pt}
\begin{tabular}{@{}c l r r r r r@{}}
\toprule
\textbf{MBS} & \textbf{Mode} & \textbf{Mean step (s)} & \textbf{Median step} (s) & \textbf{Tokens/s} & \textbf{Peak alloc. (GiB)} & \textbf{Peak HBM (MiB)} \\
\midrule
2 & Production & \textbf{6.070} & \textbf{6.021} & \textbf{21{,}593} & 36.649 & 43{,}683 \\
2 & Validation & 6.244 ($+2.873\%$) & 6.178 & 20{,}990 ($-2.792\%$) & 36.649 & 43{,}683 \\
\midrule
1 & Production & \textbf{9.823} & \textbf{9.800} & \textbf{13{,}344} & 24.152 & 29{,}911 \\
1 & Validation & 10.506 ($+6.960\%$) & 10.562 & 12{,}475 ($-6.508\%$) & 24.152 & 30{,}000 \\
\bottomrule
\end{tabular}}
\end{table*}

\subsection{Dual-Mode Distributed-Tensor Accuracy and Performance}
\label{sec:eval-dtensor}

\paragraph{Protocol.} Experiment A (Table~\ref{tab:setup}) trains
Qwen3-30B-A3B on the same 16 devices in two matched mode pairs, one with
micro-batch size (MBS) 2 and one with MBS 1.  Each run uses the same global
batch size of 32 and completes 60 steps from the same code and data.  We
compare production mode (\texttt{validate\_placement=false}), which
materializes plain local tensors when applying the plan, against validation
mode (\texttt{validate\_placement=true}), which retains DTensors for
layout propagation and boundary checking.  We report the mean over steps
2--60 after excluding profiler-export step 13 ($n=58$ per run); the fixed
workload contains 131{,}072 tokens per optimizer step, so throughput is
derived from step time.

\paragraph{Production vs. Validation.} Production is faster in both
micro-batch regimes (Table~\ref{tab:dual-mode}).  At MBS 2, validation
increases mean step time from 6.070 to 6.244 s ($+2.873\%$) and reduces
throughput by 2.792\%.  At MBS 1, validation increases step time from
9.823 to 10.506 s ($+6.960\%$) and reduces throughput by 6.508\%.  Thus,
when the smaller micro-batch makes host dispatch and operation submission
more prominent, the absolute production--validation gap grows from 0.174
to 0.684 s (3.92$\times$), and the relative step-time gap grows from
2.873\% to 6.960\%.  The profiler provides direct supporting evidence:
relative to production, validation adds 23{,}713 CPU operator events at
MBS 2 and 47{,}431 at MBS 1 %---almost exactly twice as many---
while adding no device kernels or collectives.  This is the expected signature of
validation's per-operator DTensor and metadata path; production removes
that steady-state dispatch path after applying the plan.

\paragraph{Cross-MBS Comparison.} Holding the global batch and parallel
mesh fixed, MBS 1 executes twice as many micro-batch forward/backward
passes per optimizer step as MBS 2.  Relative to MBS 2, its mean step time
is 61.8\% higher in production (9.823 vs.\ 6.070 s), but 68.3\% higher in
validation (10.506 vs.\ 6.244 s).  The corresponding throughput losses
are 38.2\% and 40.6\%, respectively.  Device activity scales in the same
direction: the profiled non-communication kernel count rises from 57{,}560
to 105{,}288 and the collective count from 9{,}554 to 18{,}854.  The
smaller MBS reduces framework peak allocation from 36.649 to 24.152 GiB,
but exposes more host-side dispatch/submission overhead; validation pays
that overhead on every operator, so its performance separates further
from production.  The clean post-profile window (steps 14\textasciitilde 60) reproduces
the trend: validation throughput is 2.597\% lower at MBS 2 and 6.638\%
lower at MBS 1.

\paragraph{Numerical and Device-Work Equivalence.} Within each MBS pair,
all 60 logged foundation-loss, total-loss, gradient-norm, and learning-rate
values match exactly at logging precision.  The rank-0 profiles also match
within each pair in non-communication kernels, collective types and counts,
and per-link transfer volume.  Framework peak allocated and reserved
memory are identical between modes; sampled HBM is identical at MBS 2 and
differs by only 89 MiB at MBS 1.  The larger validation gap is therefore
consistent with its extra host execution path, rather than a change in
accelerator work, framework memory footprint, or training numerics.  This
60-step result complements the one-step gradient-equivalence harness
(\S\ref{sec:gradequiv}); it is not a
long-horizon convergence study, which \S\ref{sec:eval-fsdp} provides
separately.

\paragraph{Measurement Scope.} Each MBS uses one production-then-validation
pair and includes a single profiled step.  Individual traces are sensitive
to communication overlap and idle-time variation, but the aggregate and
clean post-profile windows agree in both configurations. Therefore, rather than 
treating the exact percentages as a hardware-independent causal constant, 
We use the results to show the measured scaling relationship. A larger mode gap
when dispatch/submission work is more prominent.  A controlled
comparison against PyTorch's native DTensor runtime remains outside the
current evidence set.

\begin{table*}[t]
\caption{Whole-system fully-sharded results. Experiment B: Qwen3-30B-A3B,
pure FSDP on 16 dies, torch\_npu 2.9.0, steady window steps 8\textasciitilde 60
($n=53$); baseline is PyTorch FSDP2. Experiment C: 7B-class dense model,
128 ranks, TP4/DP32, PyTorch 2.6.0, steady window steps 100\textasciitilde 999
($n=900$); baseline is Megatron DDP with distributed optimizer, and its
memory rows are maxima over the 32 monitored ranks (of 128). Experiment B
throughput is nominal, derived from the fixed 131{,}072-token step shape
(the raw token counters read zero in both logs). Profiler rows are
single-step rank-0 measurements of experiment B. Disclosed configuration
deltas are listed in the text.}
\label{tab:fsdp}
{\scriptsize
\setlength{\tabcolsep}{2.5pt}
\begin{tabular}{@{}>{\RaggedRight\arraybackslash}p{3.35cm}rrrr@{}}
\toprule
& \multicolumn{2}{c}{\textbf{B: 16-die pure FSDP}} & \multicolumn{2}{c}{\textbf{C: 128-rank dense}} \\
\cmidrule(lr){2-3}\cmidrule(lr){4-5}
\textbf{Metric} & \textbf{PyTorch FSDP2} & \textbf{\sys} & \textbf{Megatron DDP} & \textbf{\sys} \\
\midrule
\textbf{Mean step time} & 3.720 s & \textbf{2.614 s} ($-$29.7\%) & 6{,}185.2$\,\pm\,$33.0 ms & \textbf{4{,}606.7$\,\pm\,$21.8 ms} ($-$25.5\%) \\
\textbf{Speedup} & 1.00$\times$ & \textbf{1.42$\times$} & 1.00$\times$ & \textbf{1.343$\times$} \\
\textbf{Throughput} & 35.23 ktok/s & \textbf{50.14 ktok/s} (+42.3\%) & 254.30 ktok/s & \textbf{341.43 ktok/s} (+34.3\%) \\
\textbf{MFU} & \multicolumn{2}{c}{not reported} & 35.65\% & \textbf{48.02\%} (+12.37 pp) \\
\textbf{Step-time CV} & \textbf{2.39\%} & 4.14\% & 0.53\% & \textbf{0.47\%} \\
\textbf{Peak allocated HBM} & 51.035 GB & \textbf{50.005 GB} ($-$1.03 GB) & 48.516 GB & \textbf{41.569 GB} ($-$14.3\%) \\
\textbf{Peak reserved HBM} & 58.145 GB & \textbf{55.201 GB} ($-$2.94 GB) & 51.793 GB & \textbf{49.150 GB} ($-$5.1\%) \\
\midrule
\multicolumn{5}{@{}l}{\emph{Experiment B, single-step rank-0 profiler:}}\\
\textbf{Stage time} & 3{,}715.1 ms & \textbf{2{,}586.7 ms} ($-$30.4\%) & & \\
\textbf{Exposed communication} & 916.7 ms & \textbf{255.4 ms} ($-$72.1\%) & & \\
\textbf{Communication overlap rate} & 70.81\% & \textbf{85.32\%} & & \\
\textbf{HCCS weighted bandwidth} & 56.78 GB/s & \textbf{125.22 GB/s} (2.21$\times$) & & \\
\bottomrule
\end{tabular}}
\end{table*}

\subsection{Topology-Aware FSDP/HSDP Comparison}
\label{sec:eval-fsdp}

This subsection reports two whole-system comparisons (experiments B and C
in Table~\ref{tab:setup}; Table~\ref{tab:fsdp}) and one baseline-free
scale demonstration (experiment D). We report whole-system differences
only; profiler observations are presented as mechanism evidence consistent
with the design rationale of \S\ref{sec:fsdp-eng}, not as per-feature
attribution.

\paragraph{Experiment B: 16-Die Comparison against PyTorch FSDP2.}
Both sides train Qwen3-30B-A3B from the same checkpoint under a pure-FSDP
mesh (shard 16; TP, EP, CP, and PP all equal to 1) with identical model,
data, and workload. Over the 53-step steady window, mean step time drops from
3.720 s to 2.614 s ($-$29.7\%), a nominal-throughput gain of 42.3\% at the
fixed step shape; peak allocated memory falls by 1.03 GB and peak reserved
by 2.94 GB; both runs reach loss 1.888 at step 60. A single-step rank-0
profiler trace corroborates the log result (stage time 3{,}715.1 ms
vs.\ 2{,}586.7 ms, $-$30.4\%) at identical logical payload (7.670 GB
all-gather, 3.817 GB reduce-scatter): exposed communication falls from
916.7 ms to 255.4 ms ($-$72.1\%) and the communication-overlap rate rises
from 70.81\% to 85.32\%. The trace is consistent with the design rationale
of \S\ref{sec:fsdp-eng}: PyTorch FSDP2's fused path surrounds its
collectives with 451.1 ms of packing/unpacking device work
(\texttt{copy\_in}/\texttt{copy\_out}), and 97 of its 125 all-gathers plus
all 49 of its reduce-scatters have buffer lengths not aligned to 512
bytes. Within PyTorch FSDP2's \emph{own} all-gathers, the non-aligned
group achieves 51.13 GB/s weighted HCCS bandwidth versus 101.96 GB/s for
the aligned group ($-$49.9\%), while \sys's non-aligned bytes are
negligible (0.0001\% of its HCCS traffic). \sys instead issues 1{,}087
all-gathers and 531 reduce-scatters per step, pays $\approx$81.7 ms of
extra per-parameter post-reduction division and an order of magnitude more
host-side launch/event time (153.2 ms vs.\ 9.5 ms of collective-API host
self time), and shows higher steady jitter (CV 4.14\% vs.\ 2.39\%). Two
configuration deltas are disclosed: prefetch depth 1 (\sys) vs.\ 0
(PyTorch), and expandable allocator segments enabled on the PyTorch side
only. The comparison is therefore between the two evaluated configurations
as whole systems: at this workload, the device-side critical-path gains
outweigh the host-side cost of per-parameter dispatch.

\paragraph{Experiment C: 128-Rank Comparison against Megatron DDP}
The second comparison uses a different baseline family at larger scale:
Megatron DDP with distributed optimizer (42 overlapped gradient buckets
over the full 32-way DP group), 128 ranks, TP4/DP32, a 7B-class dense
model, 1{,}000 steps. \sys runs as 16-way shard $\times$ 2-replica HSDP
with reshard-after-forward and prefetch depth 2. Mean step time drops from
6{,}185.2 ms to 4{,}606.7 ms ($-$25.5\%, 1.343$\times$), cluster
throughput rises from 254.30 to 341.43 ktok/s (+34.3\%), and MFU from
35.65\% to 48.02\% (+12.37 pp, same per-chip peak denominator).
Worst-observed peak allocated memory falls from 48.516 GB to 41.569 GB
($-$14.3\%). Disclosed deltas: the \sys run pads the vocabulary to
153{,}600 vs.\ 151{,}552 (+0.19\% parameters, on the \sys side),
gradient-accumulation fusion is enabled on the Megatron side only, memory
is monitored on 32 of 128 ranks, and each system was run once, so
cross-run cluster noise is not characterized.

\paragraph{Long-Horizon Numerical Alignment.} Experiment C doubles as the
paper's long-horizon numerical evidence. Over all 1{,}000 aligned steps,
the four MTP-head training losses of the two systems correlate at Pearson
$r > 0.999997$ with stepwise mean absolute differences of
0.00203--0.00313; neither run records a NaN, a skipped step, or a
loss-scale change; and final-step losses differ by at most 0.0015
(relative difference $\leq 0.034\%$). Gradient norms correlate at
$r=0.9478$ with alternating-sign differences. The two implementations are
not bit-identical because different sharding changes the FP32 reduction
order; however, no convergence-speed or stability degradation is observed.
These losses cover the four MTP heads of this model only; no downstream
evaluation is included.

\paragraph{Experiment D: 505B-Parameter Scale Demonstration, No Baseline.}
A 505B-parameter MoE (51 layers, MLA, 384 experts, top-8) trains on 384
physical Ascend 910C cards (768 ranks) with TP4/EP16/DP192, non-expert
state sharded 32-way with 6 replicas and expert state sharded 6-way with 2
replicas. The steady window (steps 30\textasciitilde 50, $n=21$) runs at 7.465 s/step
with a 0.75\% coefficient of variation, 421.4k tokens/s (36.41B
tokens/day), 67.43 TFLOP/s per logical die, and 19.1\% MFU under the run's
logging accounting. The breakdown recorded for this configuration
attributes 65.1\% of step time to compute, 31.4\% to exposed
communication, and 3.5\% to idle; FSDP communication accounts for only
2.9\% of step time, while expert-parallel token dispatch (19.5\%) and
optimizer communication (7.8\%) dominate the communication critical path.
At this scale the fully-sharded layer's own communication is a minor cost
and the optimization pressure moves to expert-parallel and optimizer
traffic. No matched baseline exists for this configuration, so we report
trainability, stability, and the achieved breakdown only, with no relative
speedup claim.

\begin{table*}[t]
\caption{Distributed-Muon comparisons at 16 NPUs (16-way fully-sharded
data parallelism, profiler step~11, rank~0; one profile per pair).
\emph{Stage} is the profiler's full-step time; equivalent-throughput
gains are $+5.74\%$, $+19.04\%$, and $+6.60\%$ respectively. The
optimizer range is Computing $+$ exposed communication $+$ free time
inside the optimizer step; the MatrixFSDP range is an approximate manual
timeline selection. Per-comparison configuration deltas are disclosed in
the text.}
\label{tab:muon}
{\small
\setlength{\tabcolsep}{3.5pt}
\begin{tabular}{@{}>{\RaggedRight\arraybackslash}p{3.1cm}>{\RaggedRight\arraybackslash}p{3.6cm}>{\RaggedRight\arraybackslash}p{1.5cm}>{\RaggedRight\arraybackslash}p{1.9cm}>{\RaggedRight\arraybackslash}p{1.9cm}>{\RaggedRight\arraybackslash}p{1.3cm}>{\RaggedRight\arraybackslash}p{2.9cm}@{}}
\toprule
\textbf{Comparison} & \textbf{Model} & \textbf{Recompute} & \textbf{Baseline stage} & \textbf{\sys stage} & \textbf{$\Delta$ stage} & \textbf{Optimizer range} (baseline $\to$ \sys) \\
\midrule
\textbf{vs.\ DMuon} & 30-layer dense Qwen3-14B-Base & full, 30/30 layers & 4{,}974.978 ms & \textbf{4{,}705.004 ms} & \textbf{$-$5.43\%} & 317.6 $\to$ 605.2 ms \\
\textbf{vs.\ MatrixFSDP} & 24-layer dense Qwen3-14B-Base & full, 24/24 layers & 4{,}548.031 ms & \textbf{3{,}820.726 ms} & \textbf{$-$15.99\%} & $\approx$394 $\to$ $\approx$488 ms \\
\textbf{vs.\ FlexShard/ DistMuon} & 48-layer MoE Qwen3-30B-A3B & full, 27/48 layers & 2{,}816.295 ms & \textbf{2{,}641.921 ms} & \textbf{$-$6.19\%} & 834.2 $\to$ \textbf{527.6} ms \\
\bottomrule
\end{tabular}}
\end{table*}

\subsection{Layout-Driven Distributed Muon Comparison}
\label{sec:eval-muon}

We compare \sys's distributed Muon against the three baseline realizations
we integrated and ran: DMuon, MatrixFSDP, and TorchTitan's
FlexShard/DistMuon. All three comparisons share one platform (16 Ascend
A3 NPUs with HCCL, PyTorch 2.6.0, CANN 8.5.2, 16-way fully-sharded data
parallelism with TP$=$CP$=$EP$=$PP$=$1, sequence length 4{,}096, global
batch 32, micro batch 2, 60 training steps) and align the Muon/AdamW
parameter split, the NS iteration count (5), the Muon learning rate
($3{\times}10^{-5}$, momentum 0.95, Nesterov), and the AdamW settings
across the two sides. Performance is measured from single-step profiler
traces (rank~0, step~11, one profile per pair) with the same
computing/exposed-communication accounting as \S\ref{sec:eval-fsdp}, not
from log step time.
Table~\ref{tab:muon} summarizes the three comparisons; the distributed
Muon of \S\ref{sect:muon} additionally trains inside experiments C and D
above, both of which optimize with Muon.

\paragraph{DMuon.} DMuon assigns each Muon matrix a global owner that
holds the full parameter, receives the FP32 gradient by reduction, and
runs Newton--Schulz locally; parameter publication and gradient reduction
are embedded as owner broadcast/reduce hooks in the forward, recompute,
and backward paths. Its optimizer range is therefore shorter than
\sys's (317.6 vs.\ 605.2 ms), but the non-optimizer path is 557.6 ms
longer (4{,}657.4 vs.\ 4{,}099.8 ms): under full-layer recompute its 212
dedicated parameters trigger 634 BF16 broadcasts (about three
materializations per parameter per step), and its total communication,
although lower (2{,}201.3 vs.\ 2{,}307.7 ms), overlaps less (78.55\%
vs.\ 85.16\%), leaving 472.1 ms exposed against \sys's 342.5 ms. The
net is $-$269.974 ms of stage time ($-$5.43\%). A disclosed secondary
delta: DMuon normalizes the NS input in FP32 and iterates in FP16, while
\sys runs the whole NS path in BF16; both sides keep FP32 optimizer
state, and this dtype-path difference does not account for the stage gap.

\paragraph{MatrixFSDP} MatrixFSDP encodes whole-matrix ownership into
the ZeRO-3 layout itself: one rank holds each Muon matrix, the others
hold empty shards, and the backward reduction delivers the full gradient
to the owner. Its optimizer range is correspondingly
communication-free and $\approx$94 ms shorter than \sys's, but the
owner broadcast/reduce traffic moves into the model path (93.47\% of its
communication time), where only 43.81\% of communication overlaps with
compute, which is significantly lower than \sys's 82.25\%, 
leaving 782.6 ms exposed against 268.4 ms. The net is $-$727.305 ms ($-$15.99\%). 
This comparison is end-to-end over the two full implementations: the baseline runs
per-matrix \texttt{torch.optim.Muon} with fixed NS coefficients while
\sys runs shape-fused batched NS with the \texttt{asym5} coefficients;
both use equivalent RMS-matching update scaling, five BF16 NS
iterations, and FP32 master weights.

\paragraph{FlexShard/DistMuon.} The DistMuon comparison controls the
fully-sharded layer itself: both sides run on \sys's FSDP with an
identical, forced parameter classification (336 Muon and 195 AdamW
parameters of the MoE model), isolating the optimizer as the variable.
DistMuon decouples storage from compute layout through a per-step
pack $\to$ all-to-all $\to$ NS $\to$ all-to-all $\to$ unpack
redistribution with a double-buffered transfer stream. Its pipeline is
effective because communication overlap is \emph{higher} than \sys's (90.64\%
vs.\ 87.22\%) and exposed communication lower, but the redistribution
adds more work than the overlap recovers: the optimizer range grows to
834.2 ms against 527.6 ms ($-36.75\%$ for \sys), and the net stage
difference is $-$174.374 ms ($-$6.19\%). Two disclosures: in this
comparison \sys's owner assignment was the simpler per-sub-batch greedy
variant, so the baseline fielded the more sophisticated (cross-bucket,
NS-cost-based) planner because the model's regular, repeated matrix shapes
limit the tail-balancing benefit such a planner can buy; and the MoE
run confirms that the FlexShard adapter consumed \sys's distributed
tensors on the native code path, not a dense-tensor fallback.

\paragraph{Issues Surfaced in the Baseline Integrations.} We report
three problems our experiments exposed in the baselines, as observed in
the pinned integration trees. (i)~\emph{DMuon, gradient lifetime under
recompute}: its custom asynchronous reduce path cleared
\texttt{param.grad} before the reduce stream had finished reading the
BF16 source gradient; under full-layer recompute the caching allocator
can reuse that storage before the enqueued cast and reduction consume
it, silently polluting the reduced gradient (observable as
\texttt{grad\_norm} or loss NaN). We fixed the integration with
\texttt{grad.record\_stream(reduce\_stream)}; the reported DMuon numbers
include the fix, and the episode is concrete evidence that embedding
optimizer communication in model hooks makes the optimizer responsible
for cross-stream tensor lifetimes that a unified shard lifecycle manages
centrally. (ii)~\emph{DMuon, memory imbalance}: owner-resident full
matrices plus FP32 master and momentum state concentrate memory on
owners (owner-resident capacities span 603M\textasciitilde 1{,}386M elements, a
2.30$\times$ max-to-min ratio); at a 40-layer configuration DMuon fails
with an HCCL memory-allocation OOM while \sys trains the same
configuration. (iii)~\emph{MatrixFSDP, coverage and allocator
footprint}: the current integration accepts only strictly
two-dimensional parameters (\texttt{torch.optim.Muon} enforces the check
at construction and at each step) and its demo gates out non-dense
models and any TP/CP/EP/PP $>$ 1, so it cannot run the fused
three-dimensional expert parameters of the MoE workload at all; and at
an equal live-tensor peak ($\approx$24.9 GB allocated on both sides) its
allocator \emph{reserved} high-water is 55.281 GB against \sys's 31.850
GB ($-42.4\%$). Structurally, both owner-resident baselines couple the
optimizer to the FSDP parameter lifecycle, i.e., broadcast/reduce hooks in
forward, recompute, and backward, whereas \sys keeps the standard shard
lifecycle and confines Muon communication to the optimizer step.

\paragraph{Expert-Weight Coverage.} Coverage is reported qualitatively
because the compared integrations expose no common per-expert numel
accounting. \sys treats a fused $[E,M,N]$ expert weight as $E$
independent $[M,N]$ matrices and orthogonalizes each expert separately
(\S\ref{sect:muon}); the MoE comparison above classified the same 336
parameters as Muon on both sides, so the fused expert weights were
exercised on the native path. MatrixFSDP is capability-limited as
described above. DMuon was evaluated on the dense model here; per the
inspected-revision analysis (\S\ref{sect:related}, Appendix~A), fused
$[E,M,N]$ weights not selected by DMuon remain on the stock FSDP2 and
original-optimizer path, and the current non-TP DMuon path would treat a
selected fused expert weight as a two-dimensional $[E,MN]$-like matrix
rather than orthogonalizing each of the $E$ experts separately.

\section{Related Work}
\label{sect:related}

\paragraph{Distributed Tensor Abstractions and Auto-Parallelism.}
GSPMD introduced annotation-driven sharding with compiler propagation for
XLA~\cite{gspmd}; OneFlow's SBP formalized placement algebra at the
framework level~\cite{oneflow}; MindSpore's auto-parallelism derives
strategies by search~\cite{mindspore}; and PyTorch's distributed tensor
brought the model to eager execution via operator-level dispatch below the
autograd engine~\cite{pytorch2,pytorchdtensor}. \sys shares the declarative surface but differs
in the interception layer: by placing sharding semantics above autograd, it
supports a precompiled-boundary execution with zero steady-state
per-operator DTensor dispatch overhead and a dual mode in which verification and production share one plan.

\paragraph{Data-Parallel Training.}
ZeRO established optimizer-state, gradient, and parameter
sharding~\cite{zero};
PyTorch FSDP productionized it with per-unit all-gather and
reduce-scatter~\cite{pytorchfsdp}, and HSDP added the replica
dimension~\cite{hsdp}. Megatron-LM composes tensor, pipeline, and data
parallelism manually~\cite{megatronlm}; MegaScale reports
10{,}000-accelerator production training~\cite{megascale}; TorchTitan
packages the stack declaratively~\cite{torchtitan}. \sys's contribution is
orthogonal to the sharding policy itself: a two-tier realization in which
intra-supernode collectives are per-parameter and zero-copy, the
cross-supernode reduction is fused without copies, and the backward
schedule contains no layer-local wait on the slow tier, and residual work
settles once at the end of the backward pass.

\paragraph{Distributed Muon.}
Muon was introduced by Jordan et al.~\cite{muon} and scaled to LLM
pretraining by Moonlight, which distributes it in bucket-based ZeRO-1
fashion~\cite{moonlight}; NorMuon improves its second-moment
normalization~\cite{normuon}. Among per-parameter realizations,
DMuon~\cite{dmuon} in the inspected revision integrates an
owner-based Muon path with the FSDP2 lifecycle: each matrix is assigned a
single owner rank holding the authoritative parameter and optimizer state,
and owner publication, gradient reduction, and prefetch run on DMuon's own
lifecycle hooks rather than those of a stock FSDP2 stack. Owner-to-all
parameter publication and all-to-owner gradient reduction are organized as
a two-stage intra-/inter-node hierarchy whose XOR owner-slot layout spreads
cross-node contention, pipelined against forward and backward compute and
published asynchronously; the owner-side Newton--Schulz iteration runs in
Gram space with symmetry-aware kernels, shape-grouped batching, and
autotuning; and owner assignment minimizes a measured makespan model.
Tensor parallelism is accommodated by a second, nested ownership level
detected from the parameters' distributed-tensor placements. Fused expert
tensors are a boundary case of the inspected revision: $[E,M,N]$ expert
weights not selected by DMuon remain on the stock FSDP2 and
original-optimizer path, while handing them to the current non-TP DMuon
path would treat them as a two-dimensional $[E,MN]$-like matrix rather
than orthogonalizing each of the $E$ experts separately; we therefore
report expert-weight coverage explicitly in the evaluation protocol
(\S\ref{sec:eval-muon}).
MatrixFSDP~\cite{matrixfsdp} takes a third path under ZeRO-3 parameter
sharding: for each 2-D weight one data-parallel rank owns the whole
matrix and the remaining ranks hold empty shards, so the ordinary
backward reduction delivers the full Muon input to the owner and the
optimizer step issues no matrix collective; a global owner planner
balances resident bytes and optimizer work, realized through owner-segment
point-to-point collectives, owner-buffer pinning, and owner-shard
checkpoint resharding. Matrices already fragmented by tensor parallelism
are excluded from owner placement and handled by the surrounding TP path.
Canzona~\cite{canzona} decouples logical optimizer assignment from
physical parameter distribution for matrix-based optimizers generally
(Muon, Shampoo, SOAP): on the data-parallel plane an $\alpha$-balanced
static partitioning assigns whole parameters to ranks so the optimizer
step is communication-free, and on the tensor-parallel plane an
asynchronous micro-group pipeline batches fragmented updates behind
compute, which is viable because tensor parallelism typically resides in the
intra-node high-bandwidth domain, in contrast to the inter-node
data-parallel plane. TorchTitan's FlexShard instead
gives the optimizer a second layout system: per-parameter compute layouts
are declared next to the DTensor storage layouts, and a general reshard
planner lowers the storage-to-compute transition onto packed all-to-all
collectives~\cite{torchtitan,flexshard}. FlexShard's abstraction is
general across optimizers; we instead trade that generality for an engine
specialized to Muon's fixed structure, keeping the algorithm on local
tensors and the communication on hardware-friendly collectives. Relative
to these systems (\S\ref{sec:muon-challenge},
Table~\ref{tab:flexshard}; versions as inspected, runtime revisions
pinned in \S\ref{sec:eval-muon}),
\sys differs along three axes: communication domains and the two-level
(shard and replica) deduplication are derived from the distribution
semantics of general multi-dimensional meshes rather than assigned per
parallel strategy or over a fixed data-parallel mesh; replica dedup
domains are aligned to machine boundaries, which is a configuration premise of
the current realization (\S\ref{sec:replica-dedup}), and \sys performs deterministic
recomputation across subgroups, so that under the canonical
mesh-to-topology mapping (Figure~\ref{fig:topo}) the evaluated
configuration sends no optimizer traffic over the slow tier; and
parameters whose matrix plane is not sharded (canonically
expert-axis-sharded MoE weights) require no Muon-specific optimizer
communication over the expert-sharding axis.

\paragraph{Supernode-Scale Systems.}
CloudMatrix384 reports production serving on a 384-die
supernode~\cite{cloudmatrix384}. The companion HyperParallel-MoE system
tackles MoE training on the same hardware class through multi-core
interleaved scheduling~\cite{hyperparallelmoe}; \sys is complementary, addressing sharding-plan execution and validation,
data-parallel communication, and the optimizer.

\section{Conclusion}
\label{sect:conclusion}

\sys makes declarative parallelization semantics explicit, verifiable, and
free of steady-state dispatch: by intercepting above autograd, one sharding
plan drives both a production mode with no per-operator DTensor dispatch
and a validation mode with
fail-fast checking, closed by gradient-equivalence testing. On top of this
sharding-plan execution and validation layer, topology-aware fully-sharded
communication and a layout-driven distributed Muon improve utilization of
the two-tier supernode network. On Atlas 900 A3 SuperPoD from 16 dies to
384 physical cards, the two tensor-runtime modes execute identical device
work and loss trajectories over the evaluated horizon
(\S\ref{sec:eval-dtensor}); the fully-sharded layer reduces mean step time
by 29.7\% against the evaluated PyTorch FSDP2 configuration at 16 dies and
by 25.5\% against Megatron DDP with distributed optimizer at 128 ranks,
with per-step losses tracking the baseline at Pearson $r>0.999997$ over
1{,}000 steps; and a 505B-parameter MoE trains at 421k tokens/s on 768
ranks with FSDP communication at 2.9\% of step time
(\S\ref{sec:eval-fsdp}); and the layout-driven distributed Muon reduces
profiler step time by 5.4--16.0\% against DMuon, MatrixFSDP, and
TorchTitan FlexShard/DistMuon integrations at 16 NPUs, where the
owner-resident baselines additionally exhibit a cross-rank memory
imbalance (OOM at a 40-layer configuration that \sys trains) and a
recompute-exposed gradient-lifetime bug (\S\ref{sec:eval-muon}).

Several limitations remain. Pipeline parallelism, though available in the
runtime, is not yet integrated into the declarative sharding plan;
context parallelism is limited to the all-gather family (no ring
attention); validation mode at LLM scale needs a lightweight reference
runner; tensor-parallel all-gather/reduce-scatter pipelining is future
work; and expert-parallel all-to-all scheduling is being addressed
separately. \sys is open source~\cite{hyperparallel}.

% === 紧接着正文输出参考文献 ===
\balance
\bibliographystyle{ACM-Reference-Format}
\bibliography{reference}

% === 换页并切换到单栏输出附录 ===
\clearpage
\onecolumn
\appendix
\section{Full Design-Contrast Matrix for Distributed Muon}
\label{app:muon-matrix}

Table~\ref{tab:flexshard-full} extends Table~\ref{tab:flexshard} to all
eleven dimensions we tracked while comparing the systems, including the
five dimensions (programming view, scheduling, fused-expert handling,
model adaptation, extensibility) that the main text omits for space. Statements about other
systems describe the versions as inspected, with runtime
revisions pinned in \S\ref{sec:eval-muon};
``not described'' marks capabilities we did not find in those revisions,
not a claim of permanent absence.

% Non-floating full-width table: the appendix is typeset in \onecolumn, so
% the matrix follows its intro text on the same page.
\begin{center}
\captionof{table}{Full design contrast of distributed Muon realizations (versions as inspected;
runtime revisions pinned in \S\ref{sec:eval-muon}): \sys's Muon-specialized static plan
against DMuon~\cite{dmuon}, MatrixFSDP~\cite{matrixfsdp},
Canzona~\cite{canzona}, and TorchTitan's general FlexShard/DistMuon
substrate~\cite{torchtitan,flexshard}.}
\label{tab:flexshard-full}
{\scriptsize
\setlength{\tabcolsep}{3pt}
\begin{tabular}{@{}>{\RaggedRight\arraybackslash}p{1.6cm}>{\RaggedRight\arraybackslash}p{3.15cm}>{\RaggedRight\arraybackslash}p{3.15cm}>{\RaggedRight\arraybackslash}p{2.75cm}>{\RaggedRight\arraybackslash}p{2.95cm}>{\RaggedRight\arraybackslash}p{2.95cm}@{}}
\toprule
 & \textbf{\sys Muon} & \textbf{DMuon} & \textbf{MatrixFSDP} & \textbf{Canzona} & \textbf{FlexShard DistMuon} \\
\midrule
\textbf{Programming view} & Local tensors; internal static global plan & FSDP2-lifecycle-integrated, owner-based Muon path with its own owner-publication, gradient-reduction, and prefetch lifecycle & ZeRO-3 parameter shards; owner placement invisible to the algorithm & Conventional optimizer interface; logical task assignment decoupled from physical distribution & Storage + compute \mbox{layouts} \\ \midrule
\textbf{Domain Derivation} & Communication domains and dedup derived from layout semantics (matrix-plane predicate) & Owner assignment over the data-parallel mesh; TP detected from placements, nested ownership & Owner-shaped ZeRO-3 shards: one rank owns each whole 2-D matrix, others hold empty shards & Logical optimizer assignment decoupled from physical parameter distribution; static whole-parameter partition (DP) & Reshard planner lowers user-declared storage-to-compute layouts \\ \midrule
\textbf{State Dedup Scope} & Shard + replica two-level; momentum stays sharded; replica state reduced to $1/R_s$ within topology-aligned subgroups & One authoritative owner per matrix (parameter + momentum); non-owners keep zero-size placeholders & Whole matrix and optimizer state on the owner rank; empty shards elsewhere; ZeRO-3-scale memory & Whole-parameter optimizer state at the statically assigned rank & Owner-style state on the compute layout (\emph{Owned}, \emph{BlockShard}) \\ \midrule
\textbf{Cross-Topology Strategy} & Replica subgroups from the local device count (contiguous rank blocks, machine-aligned under the standard rank mapping; a configuration premise, \S\ref{sec:replica-dedup}); deterministic recomputation across subgroups; no slow-tier optimizer traffic under the canonical mapping (Fig.~\ref{fig:topo}) & Two-stage intra-/inter-node hierarchy; XOR owner-slot layout spreads contention; spans both tiers & Optimizer step local; the backward reduction delivering the full gradient spans the DP mesh & DP optimizer step zero-communication; TP reconstruction confined to the intra-node high-bandwidth domain & Packed all-to-all; topology expressed through the user-chosen mesh and compute layouts; no automatic machine-boundary policy described \\ \midrule
\textbf{Batching / Load Balance} & Shape-fused batched NS; greedy owner balancing; ${\le}1$ batch in flight & Shape-grouped batched Gram-NS with SYRK kernels and autotuning; measured makespan (MILP) owner assignment & Global owner planner balancing resident bytes and optimizer work (greedy / scope-greedy / cost-aware) & $\alpha$-balanced greedy LPT (DP); micro-group balanced scheduling with greedy rollback (TP) & \emph{BucketConfig} groups and orders parameters for packed redistribution and communication-compute overlap \\ \midrule
\textbf{Scheduling} & Static plan replayed unchanged per step & Runtime hook pipeline: lookahead publication, asynchronous publish & Static global owner plan; block-local execution & Precomputed static partition map; asynchronous task execution with compute-compute overlap & Layout lowering + reshard planner \\ \midrule
\textbf{Dominant \mbox{Collective}} & Fused all-gather + relay \mbox{broadcast} & Two-stage owner-to-all broadcast / all-to-owner reduce & None in the optimizer step (the backward reduce-scatter lands the input on the owner); owner-segment P2P & DP: none; TP: fused all-to-all over micro-groups & Packed all-to-all \\ \midrule
\textbf{Mesh Coverage} & General multi-dimensional meshes; no Muon-specific optimizer communication over the expert-sharding axis when the matrix plane is unsharded & FSDP/HSDP data-parallel mesh + nested TP ownership & ZeRO-3 data parallelism; TP-fragmented matrices excluded from owner placement & ZeRO-1 DP + TP; supports Muon, Shampoo, SOAP & Layouts on named DeviceMesh axes; DistMuon constrains flat matrix-batch compute to \emph{BlockShard} on at most one non-unit mesh axis \\ \midrule
\textbf{Fused-Expert Handling} & Fused $[E,M,N]$ treated as $E$ independent $[M,N]$ matrices; per-expert NS; never flattened to $[E, MN]$ & In the inspected revision: unselected fused $[E,M,N]$ weights stay on the stock FSDP2 + original-optimizer path; the current non-TP path would treat them as a 2-D $[E,MN]$-like matrix, not $E$ per-expert $[M,N]$ matrices & Not described & Not described & Native 3-D \emph{Shard(0)} batches as listed above \\ \midrule
\textbf{Model Adaptation} & Split/cat/restore in the model adapter (fused QKV, MLA, MoE FC1) & Per-layer slices of the host stack; no model-specific transform layer & Non-matrix tensors packed into tail owners and left on AdamW & Parameter-granularity atomicity; no model-specific transform layer described & \emph{BlockShard} equal blocks, \emph{Owned} whole-matrix, or native 3-D \emph{Shard(0)} batches \\ \midrule
\textbf{Extensibility} & Per-stage hooks within Muon & Ownership-strategy plug-in; Muon-specific & Matrix optimizers of the Muon family under ZeRO-3 & Unified across matrix-based optimizers & General across optimizers \\
\bottomrule
\end{tabular}}
\end{center}

\end{document}